\documentclass[12pt]{JHEP} 
\usepackage{graphicx}
\usepackage{amssymb}
\newcommand\fverb{\setbox\pippobox=\hbox\bgroup\verb}
\newcommand\fverbdo{\egroup\medskip\noindent%
			\fbox{\unhbox\pippobox}\ }
\newcommand\fverbit{\egroup\item[\fbox{\unhbox\pippobox}]}
\newbox\pippobox

\def\ol{\overline}

\def\del{\partial}

\def\gap#1{\vspace{#1 ex}}

\def\eps{{\epsilon}}

\def\be{\begin{equation}}
\def\ee{\end{equation}}
\def\ba{\begin{array}{l}}
\def\ea{\end{array}}
\def\bea{\begin{eqnarray}}
\def\eea{\end{eqnarray}}
\def\beas{\begin{eqnarray*}}
\def\eeas{\end{eqnarray*}}
\def\eq#1{(\ref{#1})}
\def\fig#1{Figure \ref{#1}}

\def\nn{\nonumber\\}

\def\gap#1{\vspace{#1 ex}}

\def\ra{\rightarrow}
\def\eps{{\epsilon}}
\def\db{$D8$-brane}
\def\dbar{$\ol{D8}$-brane}

\title{Tachyon condensation and quark mass in modified Sakai-Sugimoto model}
\author{Avinash Dhar$^{\star~\diamond}$ and Partha Nag$^\star$ \\  
$^\star$Tata Institute of Fundamental Research, Homi Bhabha Road, \\
Mumbai 400 005, India \\~\\

$^\diamond$High Energy Accelerator Research Organization (KEK)\\
Tsukuba, Ibaraki 305-0801, Japan
\\~\\
\email{adhar@theory.tifr.res.in, parthanag@theory.tifr.res.in}}

\preprint{TIFR/TH/08-16 \\ KEK-TH-1249}

\abstract
{This paper continues the investigation of the modified Sakai-Sugimoto
model proposed in arXiv:0708.3233. Here we discuss in detail numerical
solutions to the classical equations for the brane profile and the
tachyon condensate. An ultraviolet cut-off turns out to be essential
because the numerical solutions tend to rapidly diverge from the
desired asymptotic solutions, beyond a sufficiently large value of the
holographic coordinate. The required cut-off is determined by the
non-normalizable part of the tachyon and is parametrically far smaller
than that dictated by consistency of a description in terms of
$10$-dimensional bulk gravity. In arXiv:0708.3233 we had argued that
the solution in which the tachyon field goes to infinity at the point
where the brane and antibrane meet has only one free parameter, which
may be taken to be the asymptotic brane-antibrane separation. Here we
present numerical evidence in favour of this observation. We also
present evidence that the non-normalizable part of the asymptotic
tachyon solution, which is identified with quark mass in the QCD-like
boundary theory, is determined by this parameter. We show that the
normalizable part of the asymptotic tachyon solution determines the
quark condensate, but this requires holographic renormalization of the
on-shell boundary brane action because of the presence of infinite
cut-off dependent terms. Our renormalization scheme gives an
exponential dependence on the cut-off to the quark mass. We also
discuss meson spectra in detail and show that the pion mass is nonzero
and satisfies the Gell-Mann$-$Oakes$-$Renner relation when a small
quark mass is switched on.}


\begin{document}  
    

\section{Introduction}

The model of Sakai and Sugimoto (SS) \cite{SS1} has been very
successful in reproducing many of the qualitative features of
non-abelian chiral symmetry breaking in QCD. In this model, the
`colour' Yang-Mills fields are provided by the massless open string
fluctuations of a stack of a large number $N_c$ of $D4$-branes, which
are extended along the four space-time directions and in addition wrap
a circle \cite{EW2}. In the strong coupling limit, this stack of
$D4$-branes has a dual description in terms of a classical gravity
theory. Flavour degrees of freedom are introduced in the probe
approximation as fermionic open string fluctuations between the colour
branes and an additional set of `flavour' branes
\cite{KK,BEEGK,KMMW,SJS1,BHMM,HN}, which are provided by pairs of $D8$
and \dbar s. In this setting, chiral symmetry breaking has a nice
geometrical picture. In the ultraviolet, chiral symmetry arises on
flavour \db s and \dbar s, which are located at well-separated points
on the circle, while they are extended along the remaining eight
spatial directions, including the holographic radial direction. Chiral
symmetry breaking in the infrared is signaled by a smooth joining of
the flavour branes and antibranes at some point in the bulk.

Despite its many qualitative and some quantitative successes
\cite{SS1,SS2,HSSY,HRYY,NSK,BLL,DY,ASY,AHJK,PS}, the SS model has some
deficiencies: (i) It does not have parameters associated with quark
mass and the chiral condensate.  On the other hand, there is a
parameter, the asymptotic separation between the flavour branes and
antibranes, which, within the SS scenario, finds no counterpart in
QCD; (ii) The SS model also ignores the open string tachyon between
flavour \db~and \dbar, which may be reasonable in the ultraviolet
where the branes and antibranes are well separated, but is not so at
the place in the bulk where the branes join. It is often argued that
in the curved background of the wrapped $D4$-branes, the geometry
forces flavour branes to join in the interior. While this is true of
flavour branes and antibranes that are well-separated asymptotically
(separation of the order of the antipodal distance), it cannot be the
reason when the separation is small and the branes and antibranes
meet far away from the central region. For small separation, the
effective radius of the direction on which the $D4$-branes are wrapped
is very large and so one would expect tachyon condensation to be the
primary reason for branes and antibranes meeting, as in the extremal
$D4$-brane metric. Since the tachyon field takes an infinitely large
value in the true ground state \footnote{For a recent review of this
subject, see \cite{AS1}.}, the perturbative stability argument given
in \cite{SS1}, valid for small fluctuations of the tachyon field near
the local minimum at the origin, does not apply.

It has recently been suggested in \cite{CKP,BSS,DN} that tachyon
condensation on a brane-antibrane system describes the
physics of chiral symmetry breaking in a better and more complete way.
If the brane and antibrane are well-separated \cite{BSS,DN} then one
also retains the nice geometric picture of the SS model for
non-abelian chiral symmetry breaking. The purpose of the present work
is to complete the investigations started in \cite{DN}. Here we give
detailed numerical solutions to the classical equations for the brane
profile and the tachyon. We show that the solution in which the
tachyon diverges at the point in the bulk where the brane and
antibrane meet has only one free parameter, which may be taken to be
the asymptotic separation between the flavour brane and the
antibrane. We present numerical evidence that the non-normalizable
part of the asymptotic tachyon solution is determined by this
parameter. Thus, by the usual dictionary of AdS/CFT
\cite{EW1,BKL,BKLT,AGMOO}, this parameter determines quark mass in the
boundary theory \cite{ST,AHJK}. The parameter for the asymptotic
brane-antibrane separation is present in the SS model also, but in
that setting it cannot be explained as a parameter in QCD. Thus this
parameter, which seems mysterious in the SS setting, finds a natural
explanation in our model. The presence of a non-normalizable part in
the tachyon solution necessitates introduction of an ultraviolet
cut-off. This is because in this case the numerical solutions tend to
rapidly diverge from the desired asymptotic solutions, beyond a
sufficiently large value of the radial coordinate, determined by the
magnitude of the non-normalizable part. This cut-off is parametrically
far smaller than the cut-off of order $N^{4/3}$ expected because of
the breakdown of description in terms of a 10-dimensional gravity
theory. Removing the cut-off, therefore, necessarily involves tuning
the non-normalizable part to zero. We discuss how this should be done
appropriately. We also discuss the chiral condensate and its
determination by the normalizable part of the asymptotic tachyon
solution. This determination is subtle for two reasons. One is the
fact that the space-time independent classical solutions are described
by a single parameter and hence the non-normalizable part of the
tachyon cannot be varied independent of the other parameters. The
resolution of this issue requires us to consider more general
solutions by incorporating space-time dependence. But for this one has
to go beyond the expansion in small space-time dependent fluctuations
around space-time independent solutions, basically because this
expansion is singular for the tachyon solution in the infrared. An
exact space-time dependent action is needed, which we derive. The
other subtlety has to do with the necessity of an ultraviolet
cut-off. To extract cut-off independent physics, we add counter-terms
to the \db~action to remove terms in the boundary action which are
divergent as the cut-off is formally allowed to go to infinity. With
an appropriate choice of the counter-terms we get a finite value for
the chiral condensate. Finally, we discuss meson spectra in detail and
show that the pion mass is nonzero in the presence of a
non-normalizable part of the tachyon and that it satisfies the
Gell-Mann$-$Oakes$-$Renner (GOR) relation when quark mass is small.

The organization of this paper is as follows. In the next section we
will briefly review the essential features of the modified SS model
with the tachyon present. This section also includes a more detailed
discussion of the cut-off and its implications than given in
\cite{DN}. In section \ref{numsol} we describe in detail the numerical 
solutions for the brane profile and the tachyon. This section also
contains a discussion of the parameters of the solutions and their
determination in terms of a single parameter, namely, the asymptotic
brane-antibrane separation. In section \ref{condensate} we discuss the
subtleties involved in deriving an expression for the chiral
condensate in terms of the parameters of the solutions.  We derive the
exact $5$-dimensional action in which the tachyon and brane-antibrane
separation fields have dependence on space-time as well as the
holographic coordinate and discuss solutions to the equations derived
from this action. We also discuss the counter-terms required to make
the chiral condensate finite as the cut-off is formally removed to
infinity. In section \ref{meson} we analyse small fluctuations around
the classical solution for the meson spectra. We show that the
existence of a massless pion is guaranteed if the non-normalizable
part of the tachyon solution vanishes. For a non vanishing
non-normalizable part of the tachyon solution, we obtain an expression
for the pion mass and derive the GOR relation for it. We end with a
summary and discussion in section \ref{discussion}. The Appendices
contain details of some calculations.

As this work was nearing completion, the works \cite{AK} and
\cite{HHLY} appeared which have discussed the problem of quark mass in SS 
model using different methods.

\section{\label{tss}Modified Sakai-Sugimoto model with tachyon}

The Yang-Mills part of the SS model is provided by the near horizon
limit of a set of $N_c$ overlapping $D4$-branes, filling the
$(3+1)$-dimensional space-time directions $x^\mu$
$(\mu=1,2,3~\rm{and}~0)$ and wrapping a circle in the $x^4$ direction
of radius $R_k$. Anti periodic boundary condition for fermions on this
circle gives masses to all fermions at the tree level (and scalars at
one-loop level) and breaks all supersymmetries. At low energies
compared to $l_s^{-1}$, the theory on the $D4$-branes is
$(4+1)$-dimensional pure Yang-Mills with 't Hooft coupling
$\lambda_5=(2 \pi)^2 g_s l_s N_c$, of length dimension. At energies
lower than the Kaluza-Klein mass scale, $R_k^{-1}$, this reduces to
pure Yang-Mills in $(3+1)$ dimensions. This is true in the weak
coupling regime, $\lambda_5 << R_k$, in which the dimensionally
transmuted scale developed in the effective Yang-Mills theory in
$(3+1)$ dimensions is much smaller than the Kaluza-Klein mass scale,
which is the high energy cut-off for the effective theory. In the
strong coupling regime, $\lambda_5 >> R_k$, in which the dual gravity
description is reliable, these two scales are similar. Therefore in
this regime there is no separation between the masses of glueballs and
Kaluza-Klein states. This is one of the reasons why the gravity regime
does not describe real QCD, but the belief is that qualitative
features of QCD like confinement and chiral symmetry breaking, which
are easy to study in the strong coupling regime using dual geometry,
survive tuning of the dimensionless parameter $\lambda_5/R_k$ to low
values.

Flavours are introduced in this setting by placing a
stack of $N_f$ overlapping \db s at the point $x^4_L$ and $N_f$
\dbar s at the point $x^4_R$ on the thermal circle. Massless open
strings between $D4$-branes and \db s, which are confined to the
$(3+1)$-dimensional space-time intersection of the branes, provide
$N_f$ left-handed flavours. Similarly, massless open strings between
$D4$-branes and \dbar s provide an equal number of right-handed
flavours, leading to a local $U(N_f)_L \times U(N_f)_R$ chiral gauge
symmetry on the flavour $D8$ and \dbar s. This chiral gauge symmetry
is seen in the boundary theory as a global chiral symmetry.

In the large $N_c$ and strong coupling limit, the appropriate
description of the wrapped $D4$-branes is given by the dual background
geometry. This background solution can be obtained from the Euclidean
type IIA sugra solution for non-extremal $D4$-branes by a wick
rotation of one of the four noncompact directions which the
$D4$-branes fill, in addition to wrapping the compact (temperature)
direction. In the near horizon limit, it is given by \cite{EW2,IMSY}
\bea
ds^2&=&\left(\frac{U}{R}\right)^{3/2}\left(\eta_{\mu \nu}dx^{\mu}dx^{\nu}+
f(U)~(dx^4)^2 \right)+ \left(\frac{R}{U}\right)^{3/2}\left(\frac{dU^2}
{f(U)}+U^2d\Omega_4^2\right), \nn
e^{\phi}&=&g_s\left(\frac{U}{R}\right)^{3/4},
\qquad \qquad F_4=\frac{2\pi N_c}{V_4}\eps_4, \qquad \qquad 
f(U)=1-\frac{U_{k}^3}{U^3}, 
\label{bgd}
\eea
where $\eta_{\mu \nu}=\rm{diag}(-1,+1,+1,+1)$ and $U_k$ is a constant
parameter of the solution \footnote{Note that $U$ has dimension of
length and is related to the energy scale $\tilde{U}$, which is kept
fixed in the decoupling limit, by $U=\tilde{U}\alpha'$.}. $R$ is
related to the $5$-d Yang-Mills coupling, $\lambda_5$, which is kept
fixed in the decoupling limit, by $R^3=\frac{\lambda_5
\alpha'}{4 \pi}$. Also, $d\Omega_4,~\eps_4$ and $V_4=8\pi^2/3$ are
respectively the line element, the volume form and the volume of a
unit $S^4$.

The above metric has a conical singularity at $U=U_k$ in the $U-x^4$
subspace which can be avoided only if $x^4$ has a specific
periodicity. This condition relates the radius of the circle in the
$x^4$ direction to the parameters of the background by
\be
R_k=\frac{2}{3} \left(\frac{R^3}{U_k}\right)^{\frac{1}{2}}
\label{kk}
\ee
For $\lambda_5 >> R_k$ the curvature is small everywhere and so the
approximation to a classical gravity background is reliable. As
discussed in \cite{IMSY}, at very large values of $U$, the string coupling
becomes large and one has to lift the background over to the
$11$-dimensional M-theory description.

\subsection{\label{tbb}Brane-antibrane pair with tachyon}

The effective field theory describing the dynamics of a
brane-antibrane pair in a background geometry \footnote{For
simplicity, we will discuss the case of a single flavour,
namely one brane-antibrane pair. Generalization to the multi-flavour
case can be done using the symmmetrized trace prescription of
\cite{AT}.} with the tachyon included has been discussed in
\cite{AS2,MG}. The simplest case occurs when the brane and antibrane
are on top of each other since in this case all the transverse scalars
are set to zero. This is the situation considered in
\cite{CKP}. However, in this configuration one loses the nice
geometrical picture of chiral symmetry breaking of the SS
model. The geometrical picture is retained in the case considered in
\cite{BSS,DN} where the brane and antibrane are separated in the
compact $x^4$ direction. This requires construction of an effective
tachyon action on a brane-antibrane pair, taking into account the
transverse scalars. Such an effective action with the brane and
antibrane separated along a noncompact direction has been proposed in
\cite{AS2,MG} \footnote{Also see \cite{MG+}.}. 
A generalization of this action to the case when the brane and
antibrane are separated along a periodic direction is not
known. However, for small separation $l(U)$ compared to the radius
$R_k$ of the circle, the action in \cite{MG} should provide a
reasonable approximation to the compact case. In the following we will
assume this to be so. Then, the effective low-energy tachyon action
for a $D8$ and \dbar~ pair for $l(U) << R_k$ is given, in the above
background, by \footnote{Strictly speaking, this action is valid only
when the brane and antibrane are separated along a noncompact
direction. However, as we shall see later, a posteriori
justification for using this action is provided by the classical
solutions for the brane-antibrane profile. In these solutions, for
small asymptotic separation, the brane and antibrane meet far away
from the central region. In this case, to a good approximation, the
factor $f(U)$ in the background metric can be set to identity, which
is equivalent to setting the radius $R_k$ to infinity.}
\bea
S &=& -\int d^9\sigma~V(T,l)e^{-\phi}
\left(\sqrt{-\rm{det}~A_L}+\sqrt{-\rm{det}~A_R}~\right), \nn
(A_{i})_{ab} &=& \left(g_{MN}-\frac{T^2l^2}{2\pi \alpha' Q}g_{M4}g_{4N}\right)
\del_ax^M_i \del_bx^N_i+2\pi \alpha' F^i_{ab} + 
\frac{1}{2Q}\biggl(2\pi \alpha'(D_a\tau (D_b\tau)^*
+(D_a\tau)^* D_b\tau) \nn
&& +il(g_{a4}+\del_ax^4_i g_{44})
(\tau (D_b\tau)^*-\tau^* D_b\tau)+il(\tau (D_a\tau)^*-\tau^* D_a\tau)
(g_{4b}-\del_bx^4_i g_{44})\biggr), \nn
\label{with-t1}
\eea
where
\be
Q=1+\frac{T^2l^2}{2\pi \alpha'}g_{44}, \quad 
D_a\tau=\del_a \tau-i(A_{L,a}-A_{R,a})\tau, \quad V(T,l)=g_s V(T)\sqrt{Q}.
\label{with-t2}
\ee
$T=|\tau|$, $i=L,R$ and we have used the fact that the background does
not depend on $x^4$. The complete action also includes terms involving
Chern-Simons (CS) couplings of the gauge fields and the tachyon to the
RR background sourced by the $D4$-branes. These will not be needed in
the following analysis and hence have not been included here.

The potential $V(\tau)$ depends only on the modulus $T$ of the complex
tachyon $\tau$. It is believed that $V(\tau)$ satisfies the following
general properties \cite{AS1}:
\begin{itemize}
\item
$V(T)$ has a maximum at $T=0$ and a minimum at $T=\infty$ where it
vanishes.
\item
The normalization of $V(T)$ is fixed by the requirement that the
vortex solution on the brane-antibrane system should produce the
correct relation between $Dp$ and $D(p-2)$-brane tensions. In the
present case this means $V(0)={\cal T}_8=1/(2\pi)^8~l_s^9~g_s$, 
the \db~tension.
\item
In flat space, the expansion of $V(T)$ around $T=0$ up to terms
quadratic in $T$ gives rise to a tachyon with mass-squared equal to
$-1/2\alpha'$.
\end{itemize}

There are several proposals for $V(T)$ which satisfy these
requirements \cite{AS1}, although no rigorous derivation exists.
Examples are (i) the potential used in \cite{KKKK,LP,LLM} for calculation
of decay of unstable D-branes in two-dimensional string theory
\be
V(T)={\cal T}_8~{\rm sech}{\sqrt \pi}T;
\label{pot1}
\ee
and (ii) the potential obtained using boundary string field theory
computation \cite{MZ,KMM,KL,TTU}
\be
V(T)={\cal T}_8~e^{-\frac{\pi}{2} T^2}.
\label{pot2}
\ee
Both these potentials satisfy the properties listed above. Note that
the asymptotic form of the potential in \eq{pot1} for large $T$ is
$\sim e^{-{\sqrt \pi} T}$. The linear growth of the exponent with $T$
should be contrasted with the quadratic growth for the potential in
\eq{pot2}. This difference will turn out to be important for the 
background tachyon solutions, which are discussed next.

We end this subsection with the following observation. It can be
easily seen that in the decoupling limit all factors of $\alpha'$
scale out of the entire action, without requiring any scaling of the
transverse scalar $l$ or the tachyon $\tau$. In fact, the entire
action can be rewritten in terms of $\lambda_5$ and $\tilde{U}$,
quantities that are kept fixed in the scaling limit. Henceforth, we
will use the convention $2\pi \alpha'=1$.

\subsection{\label{bgd-t}Classical equations for brane profile and tachyon}

We will now look for an appropriate classical solution of the
brane-antibrane-tachyon system. Let us set the gauge fields and all
but the derivatives with respect to $U$ of $T$ and $x_i^4$ to
zero. Moreover, we choose $x_L^4=l/2$ and $x_R^4=-l/2$ so that the
separation between the brane and antibrane is $l$. In this case, in
the static gauge the action
\eq{with-t1} simplifies to \footnote{The CS term in the action does
not contribute for such configurations.}
\be
S=-V_4 \int d^4x \int dU~V(T) \left(\frac{U}{R}\right)^{-3/4} U^4
\left(\sqrt{D_{L,T}}+\sqrt{D_{R,T}}~\right), 
\label{bgd-t1}
\ee
where $D_{L,T}=D_{R,T} \equiv D_T$ and
\be
D_T=f(U)^{-1}\left(\frac{U}{R}\right)^{-3/2}+f(U)
\left(\frac{U}{R}\right)^{3/2}\frac{{l'(U)}^2}{4}+{T'(U)}^2+T(U)^2l(U)^2.
\label{bgd-t2}
\ee
It is convenient to remove the dependence on $R$ (except for an
overall factor in the action) through a redefinition of variables,
\be
U=u/R^3, \quad l(U)=R^3 h(u), \quad U_k=u_k/R^3.
\label{redef}
\ee
In terms of the new variables, we get
\be
S=-V_4 R^{-9} \int d^4x \int du~u^{13/4}~V(T)
\left(\sqrt{d_{L,T}}+\sqrt{d_{R,T}}~\right), 
\label{new-t1}
\ee
where 
\be
d_{L,T}=d_{R,T} \equiv d_T=f(u)^{-1} u^{-3/2}+f(u)~u^{3/2}\frac{{h'(u)}^2}{4}+
{T'(u)}^2+T(u)^2 h(u)^2,
\label{new-t2}
\ee
with $f(u)=(1-u_k^3/u^3)$. 

\begin{figure}[htb] 
\centering 
\includegraphics[height=5cm,width=9cm]{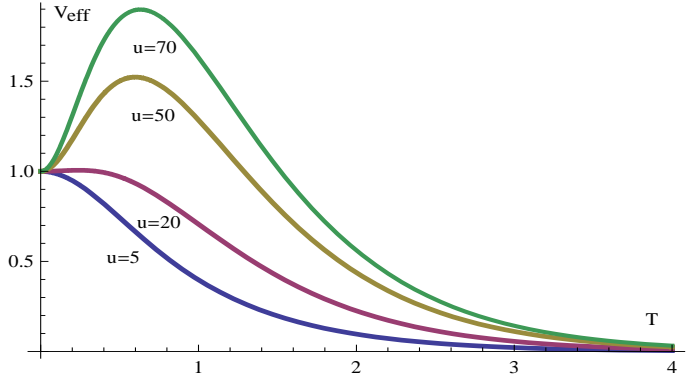} 
\caption{The effective potential $V_{\rm eff}$ as a function of $T$ for
different values of $u$ for a fixed non-zero value of $h$.}  
\label{fig.0}
\end{figure}

The effective potential for the tachyon can be obtained from this action 
by setting $T'=h'=0$. It is 
\be
V_{\rm eff}(T,l) \sim {\rm sech}{\sqrt \pi}T \sqrt{1+u^{3/2}T^2h^2}
\label{eff-pot}
\ee
In \fig{fig.0} we have plotted $V_{\rm eff}$ as a function of $T$ for
various values of $u$. We see that a perurbatively stable minimum at
$T=0$ for large values of $u$ turns into an unstable maximum at a
sufficiently small value of $u$. This is true for any fixed, non-zero
value of $h$. Moreover, the value of $u$ at which there is an unstable
maximum at $T=0$ increases as $h$ decreases.

The equations of motion obtained from the action \eq{new-t1} are
\bea
\left(\frac{u^{\frac{13}{4}}}{\sqrt{d_T}} T'(u)\right)' 
&=& \frac{u^{\frac{13}{4}}}{\sqrt{d_T}} 
\left[T(u)h(u)^2+\frac{V'(T)}{V(T)}(d_T-T'(u)^2)\right],
\label{eq-t} \\
\left(\frac{u^{\frac{13}{4}}}{\sqrt{d_T}}\frac{f(u)}{4}
u^{\frac{3}{2}}h'(u)\right)' 
&=& \frac{u^{\frac{13}{4}}}{\sqrt{d_T}}
\left[T(u)^2 h(u)-\frac{V'(T)}{V(T)}\frac{f(u)}{4}
u^{\frac{3}{2}}h'(u) T'(u)\right]. \nn
\label{eq-l}
\eea
Note that the `prime' on $V(T)$ denotes a derivative w.r.t. its
argument $T$ and not a derivative w.r.t. $u$.

This is a complicated set of coupled nonlinear differential equations
which can be solved completely only numerically. To get some insight
into the kind of solutions that are possible, however, we had analysed
these equations in \cite{DN} for large $u$ and for $u$ near the
brane-antibrane joining point in the bulk. For these values of $u$ the
equations simplify and can be treated analytically. For the sake of
completeness, we will summarize the results of this analysis here
before proceeding to describe numerical solutions to these
equations. As in the case without the tachyon, we are looking for
solutions in which the brane and antibrane have a given asymptotic
separation $h_0$, i.e. $h(u) \ra h_0$ as $u \ra\infty$, and they join
at some interior point in the bulk, i.e. $h(u) \ra 0$ at $u=u_0 \geq
u_k$ \footnote{The inequality results from the lower bound on
$u$.}. Moreover, we want the tachyon (i) to vanish as $u \ra \infty$
so that the chiral symmetry is intact in the ultraviolet region and
(ii) to go to infinity as $u$ approaches $u_0$ so that the QCD chiral
anomalies are reproduced correctly \cite{CKP}.

\subsection{\label{large-u}Solution for large $u$}

Here we seek a solution in which $h(u)$ approaches a constant $h_0$
and $T$ becomes small as $u \ra \infty$. For small $T$ one can
approximate $V'/V \sim -\pi T$
\footnote{This follows from the general properties of the potential
discussed in section \ref{tbb}.}. If $T$ and $h'$ go to zero
sufficiently fast as $u \ra \infty$ such that to the leading order one
might approximate $d_T \sim u^{-3/2}$, then \eq{eq-t} can be
approximated to
\be
\left(u^4~T'(u)\right)'=h_0^2~u^4~T.
\label{asym-t}
\ee
The general solution of this equation is
\be
T(u)=\frac{1}{u^2}(T_+ e^{-h_0 u}+T_- e^{h_0 u}).
\label{sol-t}
\ee
In writing this solution we have ignored a higher order term in $1/u$
for consistency with other terms in equation \eq{eq-t} that we have
neglected at large $u$. We will discuss consistency of this solution
below. Let us first discuss the solution for $h(u)$.

The fact that the tachyon takes small values for large $u$ makes it
irrelevant for the leading asymptotic behaviour of $h$, which can be
extracted from \eq{eq-l} by setting the r.h.s. to zero. The resulting
equation is
\be
\left(u^{\frac{11}{2}}h'(u)\right)'=0,
\label{asym-l}
\ee
which has the solution
\be
h(u)=h_0-h_1 u^{-9/2}.
\label{sol-l}
\ee
Here $h_1$ is restricted to positive values so that the branes come
together in the bulk. For SS model without the tachyon,
$h_1=\frac{4}{9} u_0^4 f_0^{1/2}$, where $f_0=f(u_0)$, $u_0$ being the
value of $u$ where the branes meet in the bulk.

It is easy to convince oneself that the only solution to equations
\eq{eq-t} and \eq{eq-l} in which $T$ vanishes asymptotically and $h$ 
goes to a constant is \eq{sol-t} with $T_-=0$. In particular, for
example, these equations have no solutions in which $T$ vanishes
asymptotically as a power law.

\subsection{\label{cutoff}Quark mass and the ultraviolet cut-off}

In the tachyon solution \eq{sol-t}, the exponentially falling part
satisfies the approximations under which \eq{asym-t} was derived for
any large value of $u$. The exponentially rising part will, however,
eventually become large and cannot be self-consistently used. This is
because for sufficiently large $u$, there is no consistent solution
for $T$ which grows exponentially or even as a power-law to the
original equations \eq{eq-t} and \eq{eq-l}, if we impose the
restriction that $h(u)$ should go to a constant asymptotically. This
puts a restriction on the value of $u$ beyond which the generic
solution \eq{sol-t} cannot be used. The most restrictive condition
comes from the approximation $d_T
\sim u^{-3/2}$. This requires the maximum value, $u_{\max}$, to
satisfy the condition
\be
T_+^2 e^{-2h_0u_{\max}}+T_-^2 e^{2h_0u_{\max}}
<< \frac{u_{\max}^{5/2}}{2 h_0^2}
\label{umax}
\ee
For generic values of $|T_\pm|$ and $h_0$, this inequality determines a
range of values of $u_{\max}$ for which the solution
\eq{sol-t} can be trusted. The value $T_-=0$ is special since in
this case there is no upper limit on $u_{\max}$, except the cut-off
that comes from the fact that the $10$-dimensional description of the
background geometry breaks down beyond some very large value ($\sim
N_c^{4/3}$) of $u$. However, as is clear from \eq{umax}, for nonzero
$|T_-|$ one needs to choose a much smaller value of
$u_{\max}$. Numerical calculations reported in the next section bear
out this expectation.

It is important to emphasize that the ultraviolet cut-off we are
talking about here does not merely play the usual role of a cut-off
needed in any example of AdS/CFT with a non-normalizable part present
in a solution to the bulk equations. The point is that there is no
growing solution to the tachyon equation in the ultraviolet which is
consistent with a brane profile that goes to a finite asymptotic
brane-antibrane separation. This constraint limits the value of $u$
up to which the asymptotic solutions, \eq{sol-t} and \eq{sol-l}, can be
trusted. 

One way to think about the inequality \eq{umax} is the
following. Suppose for given values of $|T_\pm|$ we have chosen the
largest value of $u_{\max}$ consistent with \eq{umax}. Increasing
$u_{\max}$ further would then be possible only if $|T_-|$ is decreased
appropriately, while $|T_+|$ can be kept fixed, as $u_{\max}$ is
increased. To be concrete, let us keep $|T_+|$ and$|T_-|
e^{h_0u_{\max}}$ fixed as $u_{\max}$ is increased. The process
of ``removing the cut-off'' can then be understood as increasing
$u_{\max}$ and simultaneous decreasing $|T_-|$ while keeping $|T_+|$
and the combination $|T_-| e^{h_0u_{\max}}$ fixed. In this process, at
some point $|T_+| e^{-h_0u_{\max}}$ would become much smaller than
$|T_-| e^{h_0u_{\max}}$. As we shall see in the next section, however,
limitations due to numerical accuracy prevent us from tuning $|T_-|$
to very small values, or equivalently tuning $u_{\max}$ to be very
large. Thus we are numerically restricted to rather small values of
$u_{\max}$. For values of $u$ larger than $u_{\max}$, the inequality
\eq{umax} breaks down and consequently the asymptotic solution \eq{sol-t} 
is not applicable. Clear evidence for this breakdown is seen in the
numerical calculations reported in the next section.

It is natural to associate $T_-$ with the quark mass since this
parameter comes with the growing solution. Evidence for this will be
given in section \ref{meson} where we will show that for a small
nonzero value of this parameter, the pion mass is nonzero and
proportional to it. It is also natural to associate $T_+$ with the
chiral condensate because it comes with the normalizable solution. It
turns out that this association too is consistent, though this part of
the story is somewhat more complicated, as we shall see in section
\ref{condensate}.

It is interesting to mention here that keeping the combination $|T_-|
e^{h_0u_{\max}}=\rho$ fixed as the cut-off becomes large implies an
exponential dependence of $|T_-|$ on the $u_{\max}$, i.e. $|T_-|=\rho
e^{-h_0u_{\max}}$. A similar dependence of the quark mass on the
cut-off has been observed in \cite{AK,HHLY}, though the methods used
for computing quark mass in these works are quite different from
ours. In \cite{HHLY} the cut-off arises from the location of a
$D6$-brane, which is additionally present in that model, thereby
giving a physical meaning to the cut-off.

\subsection{\label{small-u}Solution for $u \sim u_0$}

Here we are looking for a solution in which $h \ra 0$ and $T \ra
\infty$ as $u \ra u_0$. Let us assume a power law ansatz, namely
\be
h(u) \sim (u-u_0)^\alpha, \quad \quad T(u) \sim (u-u_0)^{-\beta}.
\label{ansatz}
\ee
For a smooth joining of the brane and antibrane at $u_0$, the
derivative of $h$ must diverge at this point, which is ensured if
$\alpha < 1$. Since for this ansatz $T'^2$ is the largest quantity for
$u \ra u_0$, we can approximate $d_T \sim T'(u)^2$. We will also need
the asymptotic form of the potential $V(T)$ for large $T$, which
depends on the specific potential being used. From the asymptotic form
of the potential in \eq{pot1}, we get $V'(T)/V(T) \sim -\sqrt{\pi}$,
while for the potential in \eq{pot2}, we get $V'(T)/V(T)
\sim -\pi T$. Putting all this in \eq{eq-t} and \eq{eq-l}, it is easy to 
verify that these equations cannot be satisfied by the ansatz
\eq{ansatz} for the potential \eq{pot2}. They are, however, satisfied
for the potential in \eq{pot1}. In fact, in this case the powers as
well as the coefficients all get fixed \footnote{In \cite{BSS} the
power of $(u-u_0)$ with which the brane-anibrane separation falls-off
in the bulk has been left undetermined. This power is actually
determined by \eq{eq-t} and \eq{eq-l}, as can be easily checked by
consistently expanding these equations on both sides and going beyond
the leading order in powers of $(u-u_0)$. We have also verified this
power by numerical calculations reported in the next section.}:
\bea
h(u) &=& \sqrt{\frac{26}{\pi u_0f_0}} u_0^{-3/4}(u-u_0)^{1/2}+ \cdots,
\label{sol-la} \\
T(u) &=& \frac{\sqrt{\pi}}{4} f_0 u_0^{3/2} (u-u_0)^{-2}
+ \cdots,
\label{sol-ta}
\eea

An important feature of the above solution is that it depends only on
a single parameter, namely the value of $u_0$. We have checked that
this feature persists in the next few higher orders in a power series
expansion in $(u-u_0)$. This is in sharp contrast to the asymptotic
solution \eq{sol-t}, \eq{sol-l} which depends on all the four expected
parameters, $T_+,~T_-,~h_0,~h_1$. This reduction in the number of
parameters is similar to what happens in the SS model where the
solution for $u \sim u_0$ depends only on one parameter, although the
asymptotic solution depends on two parameters. In the present case the
reduction in the number of parameters is even more severe; the
solution for $u \sim u_0$ matches with only a one-parameter subspace of
the four-parameter space of asymptotic solutions. As we will discuss
later, this one-parameter freedom of the classical solution turns out
to be analogous to the freedom to add a bare quark mass in QCD.

For completeness, we note that there exists another solution in which
$T$ does not diverge but goes to a nonzero constant as
$u \ra u_0$. In this case we can approximate $d_T \sim
f(u)u^{3/2}{h'(u)}^2/4$. Substituting in \eq{eq-t} we see that the
l.h.s. diverges as $(u-u_0)^{-\alpha}$. The first term on the
r.h.s. vanishes as a positive power, but the second term diverges as
$(u-u_0)^{\alpha-1}$, since $\alpha < 1$. For consistency we must have 
$\alpha=1/2$. The resulting solution
\bea
h(u) &=& \frac{4}{u_0} (f_0(5f_0+3))^{-1/2} (u-u_0)^{1/2}+ \cdots, 
\label{sol-lb} \\
T(u) &=& t_0+\frac{2 u_0^{-1/2}}{(5f_0+3)} \frac{V'(u_0)}{V(u_0)}(u-u_0)+ 
\cdots,
\label{sol-tb}
\eea
also satisfies \eq{eq-l}. Note that no special condition was required
for the tachyon potential to get this solution; this solution exists
for any potential.

\section{\label{numsol}Numerical solutions}

The equations \eq{eq-t}, \eq{eq-l} cannot be solved analytically. One
needs to use numerical tools to get a solution. We have made use of
mathematica for this. Also, for numerical calculations we have chosen
the potential \eq{pot1}, since there is no diverging solution for
$T(u)$ for $u \sim u_0$ for the potential \eq{pot2}, as discussed above.

The numerical calculations are easier to do if we start from the
$u=u_0$ end and evolve towards the large $u$ end. This avoids the
fine-tuning one would have to do if one were to start from large
values of $u$, where the general solution has four parameters, and end
on a one-parameter subspace for $u \sim u_0$. We must also satisfy the
requirement of working in the parameter region of the background
geometry corresponding to the strong coupling.  In addition, we need
to ensure that the asymptotic separation between flavour branes and
antibranes is small compared to the radius of the $x^4$
circle. Mathematically, these requirements are $\lambda_5=8\pi^2 R^3
\gg 2\pi R_k$ and $l_0 \ll \pi R_k$. Using \eq{kk} and \eq{redef}, one
gets $R^3=\frac{3}{2}R_k\sqrt{u_k}$. Then, these requirements become
$\frac{1}{36\pi^2} \ll u_k \ll \frac{4\pi^2}{9h_0^2}$. Throughout our
numerical calculations we will work with $u_k=1$, which satisfies the
first condition easily, while it requires from the second that $h_0
\ll \frac{2\pi}{3}$. This condition is also easily satisfied by
choosing $u_0 \gg u_k=1$ \footnote{As we shall see below, the
asymptotic separation decreases with increasing value of $u_0$, as is
the case for the SS model.}. For such values of $u_0$, $f(u) \sim 1$
for all $u \geq u_0$.

\begin{figure}[htb] 
\centering 
\includegraphics[height=5cm,
width=12cm]{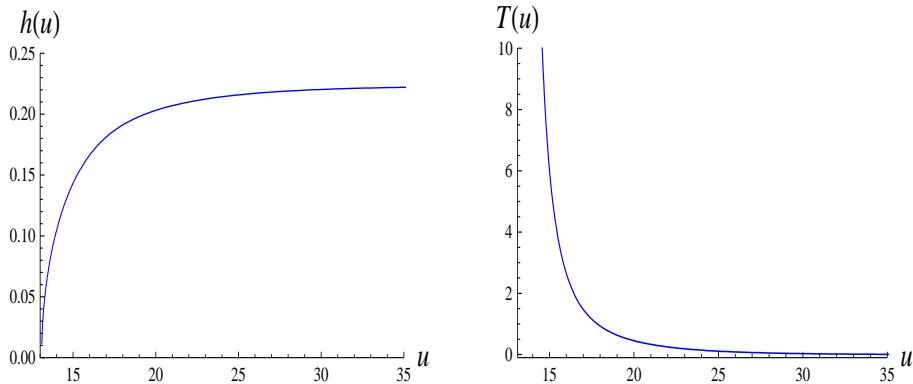}
\caption{The brane profile and the tachyon solution for $u_0=12.7$.}
\label{fig.1}
\end{figure}

The boundary conditions are imposed using \eq{sol-la}, \eq{sol-ta} at
a point $u=u_1$ which we choose as close to $u_0$ as allowed by
numerics. Generally we were able to reduce $(u_1-u_0)$ down to about
$0.1$ percent of the value of $u_0$. Starting from the values of
$T(u_1),~T'(u_1),~h(u_1)$ and $h'(u_1)$ obtained from \eq{sol-la},
\eq{sol-ta} at $u=u_1$, the system was allowed to evolve to larger
values of $u$. \fig{fig.1} shows an example for $u_0=12.7$. Solutions for
both $h(u)$ and $T(u)$ are shown.

\subsection{Verification of the UV and IR analytic solutions}

\begin{figure}[htb] 
\centering 
\includegraphics[height=5cm,
width=12cm]{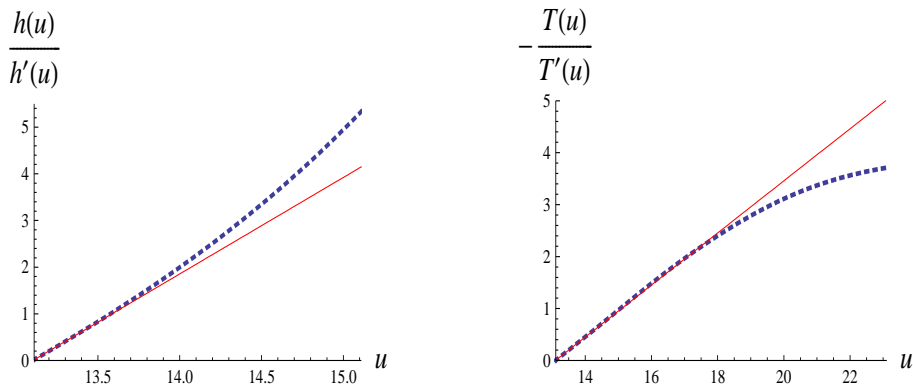}
\caption{Numerical verification of exponents in the IR behaviour of 
brane profile and tachyon. The fits give the two exponents
respectively to be $0.50$ and $-2.07$ for $u=13.1$.}
\label{fig.2}
\end{figure}

\begin{figure}[htb] 
\centering 
\includegraphics[height=5cm,
width=12cm]{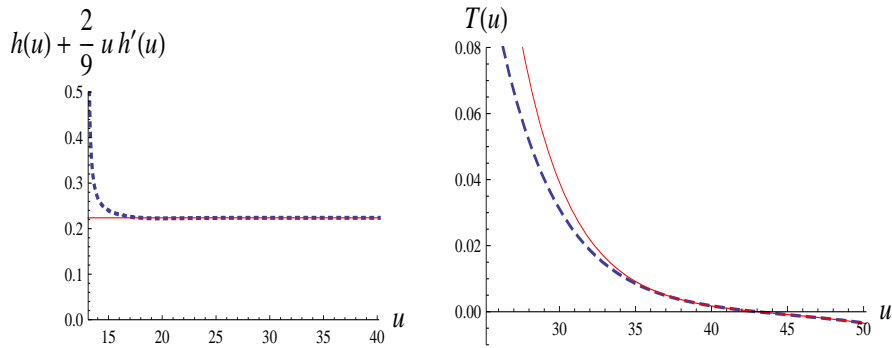}
\caption{Numerical verification of the asymptotic form of the brane 
profile and the tachyon.}  
\label{fig.3}
\end{figure}

From the numerical solutions one can verify that $h(u)$ and $T(u)$ are
given by the forms \eq{sol-la}, \eq{sol-ta}, for $u \sim
u_0$. \fig{fig.2} shows the impressive fits between the numerical data
and the analytical expectations for the powers of $(u-u_0)$ for $h(u)$
and $T(u)$. We have plotted $h(u)/h'(u)$ and $T(u)/T'(u)$, calculated
from the numerical solutions, as functions of $u$. The numerical data
are plotted in dashed lines while the theoretical solutions are
plotted in solid lines. As one can see, these graphs are linear at the
IR end and their slopes turn out to be close to the expected values
$0.5$ and $-2$ respectively. In fact, the numerical and the theoretical
curves entirely overlap in the IR region of $u$, as shown in
\fig{fig.2}.
At the other end also, namely for large $u$, one can verify that the
numerical solutions have the analytic forms
\eq{sol-l}, \eq{sol-t}. The goodness of the fits of these analytic 
forms to numerical data is shown in \fig{fig.3} where again the two curves
overlap in the asymptotic region of $u$. The fits yield values of the
four parameters:
$h_0=0.224,~h_1=-16068,~T_+=29194.5,~T_-=-1.25 \times 10^{-4}$ for
$u_0=13.1$.

\subsection{Behaviour of the non-normalizable part}

\begin{figure}[htb] 
\centering 
\includegraphics[height=5cm,
width=12cm]{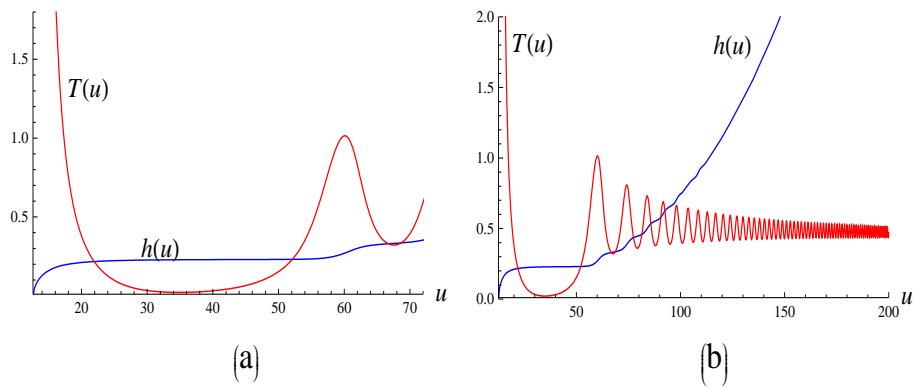}
\caption{Solutions for two different large values of $u$.}  
\label{fig.4}
\end{figure}

For $T_- \neq 0$, extending numerical calculations much beyond the
values of $u$ shown in \fig{fig.1} meets with a difficulty. It turns
out that for small $u_0$, $T_-$ is positive. Since $T_-$ is the
coefficient of the rising exponential in $T(u)$, for a sufficiently
large value of $u$ this term dominates and so $T(u)$ begins to rise
\footnote{We would like to thank Matt Headrick for a discussion on
this point and some other aspects of our numerical
calculations.}. Eventually, $T$ becomes so large that the conditions
under which the asymptotic solutions \eq{sol-l}, \eq{sol-t} were
obtained no longer apply. \fig{fig.4} illustrates this; it shows the
solutions for $u_0=12.7$ for two different large values of $u$.
In Figure 5(a), after falling very fast, $T$ rises and then falls
again. Almost simultaneous with this is a rapid rise of $h$ from one
nearly constant value to a higher constant value. Evidently, this
behaviour continues indefinitely with $u$, as can be seen in Figure
5(b) \footnote{In \cite{BSS}, the authors claim that this effect is
due to sensitivity of the solutions to the boundary conditions at the
infrared end at $u=u_1$, which must necessarily be chosen slightly
away from the actual value $u_0$. We have not found any evidence for
this sensitivity. On the other hand, it is clear that the
approximation made in deriving the asymptotic solution, \eq{sol-t},
\eq{sol-l}, must break down for sufficiently large $u$, for any
non-zero value of $T_-$. We see convincing numerical evidence for
this. Further evidence of this follows.}.

\begin{figure}[htb] 
\centering 
\includegraphics[height=4.2cm,
width=12cm]{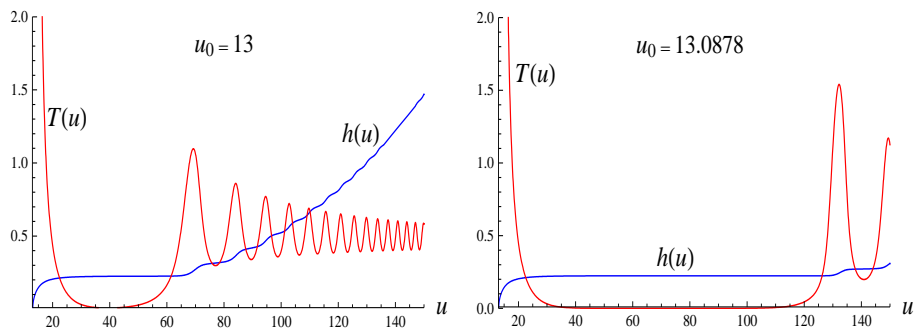}
\caption{Numerical solutions for increasing values of $u_0$
for positive $T_-$.}
\label{fig.5}
\end{figure}

\begin{figure}[htb] 
\centering 
\includegraphics[height=5cm,
width=12cm]{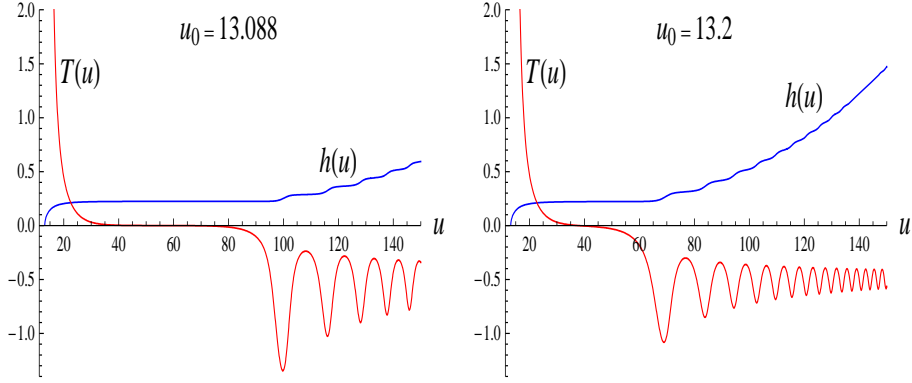}
\caption{Numerical solutions for increasing values of $u_0$ 
for negative $T_-$.}  
\label{fig.6}
\end{figure}

\begin{figure}[htb] 
\centering 
\includegraphics[height=5cm,
width=8cm]{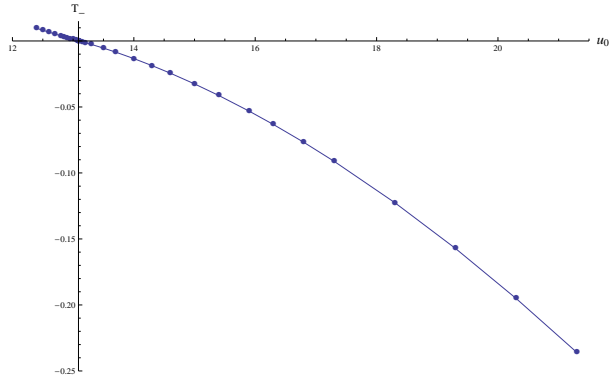}
\caption{$T_-$ as a function of $u_0$.}  
\label{fig.7}
\end{figure}

The value of $T_-$ decreases with increasing $u_0$. This can be easily
deduced from the fact that the maximum value of $u$ up to which the
asymptotic solutions \eq{sol-t}, \eq{sol-l} apply, namely before the
oscillations begin, increases with increasing $u_0$. \fig{fig.5} illustrates
this by showing the solutions for increasing values of $u_0$, close to
where $T_-$ is small. As one can see, increasing the value of $u_0$ by
a very small amount, from $u_0=13$ to $u_0=13.0878$, dramatically
increases the threshold for oscillatory behaviour of $T$ from $u \sim
50$ to $u \sim 120$!
As $u_0$ increases further, $T_-$ decreases, becomes zero
\footnote{We have found that $T_-=1.92 \times 10^{-9}$ at 
$u_0 \sim 13.0877781$. Fine-tuning $u_0$ such that $T_-$ is precisely
zero is hard. This requires numerical methods which are beyond the
scope of those used here. However, the trend is clear from \fig{fig.5} and
\fig{fig.6}.} and eventually negative. Since we want to interpret $T_-$ as the
bare quark mass parameter, negative values for it are
allowed. However, a large value for $|T_-|$ will eventually again make
$T$ large in magnitude for large enough $u$. So once again we expect
that at some sufficiently large $u$, $T$ will become so large that the
conditions under which the asymptotic solutions \eq{sol-t}, \eq{sol-l}
were obtained no longer apply. So, as before, one should find
oscillations in $T(u)$, which now start at smaller and smaller $u$ as
$u_0$ grows. This is indeed seen to be the case, as is evident in
\fig{fig.6}.
This happens because $|T_-|$ grows with $u_0$, beyond the value at
which it becomes zero. \fig{fig.7} shows the change of $T_-$ with $u_0$.
We see that $T_-$ vanishes at $u_0 \sim 13.0878$ and $|T_-|$ grows on
both sides away from this value. It is hard to understand what is
special about this value of $u_0$. One might have thought that the
role of zero mass would be played by the antipodal configuration,
which has $u_0=u_k$, and is beyond our approximation. It is possible
that this is an artifact of using the approximate action,
(\ref{with-t1}), valid for a noncompact $x^4$ coordinate, although the
value $u_0 \sim 13.0878$ is fairly large and seems to be within the
validity of our approximation. We also note that for negative $T_-$,
negative $T(u)$ can be avoided by imposing a suitable cut-off on
$u$. As we have already discussed, the cut-off is in any case required
to fulfil the condition \eq{umax} so that the asymptotic solutions
\eq{sol-t}, \eq{sol-l} may apply.

\subsection{Behaviour of the asymptotic brane-antibrane separation}

\begin{figure}[htb] 
\centering 
\includegraphics[height=5cm,
width=8cm]{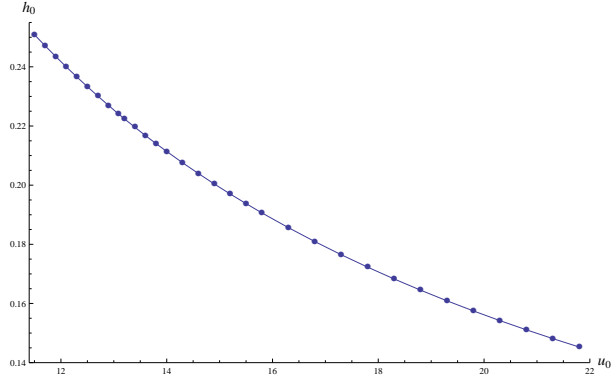}
\caption{$h_0$ as a function of $u_0$.}  
\label{fig.8}
\end{figure}

Another interesting quantity is the asymptotic brane-antibrane
separation, $h_0$, as a function of $u_0$. This quantity has been
plotted in \fig{fig.8}.
We see that $h_0$ steadily decreases through the special value $u_0
\sim 13.0878$. Although we do not have an analytical formula for the 
dependence of $h_0$ on $u_0$ for large values of the latter, the trend
in \fig{fig.8} seems to indicate that it decreases to zero as $u_0$ becomes
large. Presumably the brane-antibrane pair overlap and disappear as
$u_0$ goes to infinity. This is consistent with the trend of
increasing bare quark mass for increasing values of $u_0$ (far beyond $u_0
\sim 13.0878$) which we have seen in \fig{fig.7}. Therefore, unlike in the
Sakai-Sugimoto model, the disappearance of the brane-antibrane pair
for $u_0=\infty$ can be understood in the present setup as the
infinite bare quark mass limit.

It should be clear from the above discussion that the limit $h_0
\rightarrow 0$ does not reduce to the case of overlapping $D8$-branes
and $\overline{D8}$-branes considered in \cite{CKP}. For this case,
one must begin afresh with $x_i^4=0,~l=0$ in the action
(\ref{with-t1}). However, the classical equation for $T$ can be
obtained from the equation (\ref{eq-t}) by setting $h=0$ in it. As
above, we find that solutions which are divergent in the IR depend on
only one free parameter. For further details about the tachyon
solutions in this case, we refer the interested reader to the Appendix
\ref{A}.

\subsection{Comparison with the Sakai-Sugimoto solution}

Finally, we must ensure that the solution with the tachyon has lower
energy compared to the SS model. The energy density in the modified
model is given by
\bea
E_{\rm T} &=& 2V_4R^9~V(0) \int_{u_0}^{u_{\rm max}} du~E_{\rm T}(u), 
\nonumber \\
E_{\rm T}(u) &=& u^{13/4}~\frac{V(T)}{V(0)} \sqrt{u^{-3/2}
+\frac{1}{4}u^{3/2}{h'(u)}^2+{T'(u)}^2+T(u)^2 h(u)^2},
\label{energy}
\eea
while for the SS model it is given by
\bea
E_{\rm SS} &=& 2V_4R^9~V(0)\int_{u_0}^{u_{\rm max}} du~E_{\rm SS}(u), 
\nonumber \\
E_{\rm SS}(u) &=& u^{13/4}\sqrt{u^{-3/2}+\frac{1}{4}u^{3/2}{h'_{\rm SS}(u)}^2}.
\label{ssenergy}
\eea
To get these expressions for energy density, we have set $f(u)$ to
unity, which is a good approximation for large $u_0$. Also, in the SS
model one must use the solution of the tachyon free equation, $h'_{\rm
SS}(u)=2u_0^4u^{-3/2}(u^8-u_0^8)^{-1/2}$.

Close to $u_0$, in the IR, the exponentially vanishing tachyon
potential suppresses contribution to $E_{\rm T}$ compared to $E_{\rm
SS}$. Since the UV solutions for the two models are almost identical
\footnote{There is a caveat here. Strictly speaking this is true only
when the coefficient of the non-normalizable term, $T_-$, in the
asymptotic tachyon solution \eq{sol-t} vanishes. As we have discussed,
when $T_-$ is nonzero, one must introduce a cut-off, $u_{\rm max}$,
chosen carefully such that the asymptotic solution is satisfied. In
particular, one must ensure $T$ is positive in the region below
$u_{\rm max}$. In the calculations reported here and earlier in this
section, this is what we have done.}, one might argue that the energy
for the modified model must be lower than that for the SS
model. However, for $u \gtrsim u_0$ there is a competition between the
exponentially vanishing tachyon potential and the power law increase
of the square-root factor coming from $|T'|$ in the integrand $E_{\rm
T}(u)$ in \eq{energy}. This results in a local maximum in $E_{\rm
T}(u)$ at some value of $u$, which can be easily estimated
analytically. The relevant quantity, $$e^{-\frac{\pi}{4}
u_0^{3/2}(u-u_0)^{-2}}(u-u_0)^{-3},$$ has a maximum at
$u=u_0+(\frac{\pi}{6})^{1/2}u_0^{3/4}$. For small $u_0$, the position
of the maximum is close to $u_0$, so in this case the argument about
the IR behaviour of the integrand in \eq{energy} is not very clean,
except in the very deep IR. But since the position of the maximum
grows with increasing $u_0$ as $u_0^{3/4}$, our argument should hold
for large values of $u_0$, which is precisely where the action for the
modified model can be trusted. However, the expression used for
estimating the position of the local maximum breaks down if it is too
far away from $u_0$. So, in practice we need to do a numerical
calculation to see what the real story is. As we will see in the
numerical plots given below, what really happens is that for
relatively large values of $u_0$ the integrand $E_{\rm T}(u)$
increases rapidly at first, then slows down almost to a constant and
finally settles into an asymptotic power law increase similar to that
of the integrand $E_{\rm SS}(u)$ for the SS model. Moreover, the place
where the rapid increase begins shifts to larger values of $u$ as
$u_0$ increases, in accordance with the above expectation.

We have numerically evaluated the integrals in \eq{energy} and
\eq{ssenergy}. Because the relation between $u_0$ and the asymptotic
brane-antibrane separation is different in the two models, a given
value of $u_0$ corresponds to two different values of the latter and
vice versa. We have chosen to do the comparison for the same value of
the asymptotic brane-antibrane separation in the two models, but the
conclusions are similar with the other choice as well. 
\begin{figure}[htb] 
\centering 
\includegraphics[height=5cm,width=14cm]{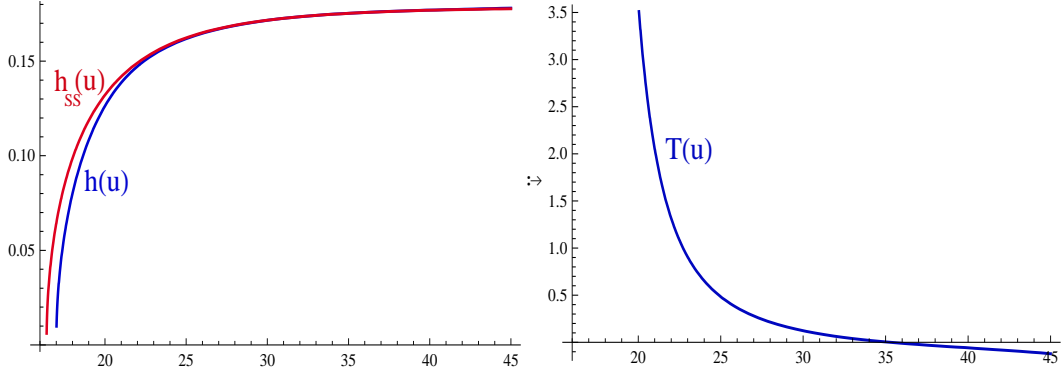}
\caption{$h(u)$ and $T(u)$ profiles for $u_0=17$. For comparison, 
$h_{\rm SS}$ profile has also been plotted after adjusting the 
value of $u_0$ to $16.4$ for it since this value of $u_0$ produces 
the same asymptotic brane-antibrane separation.}
\label{fig.9}
\end{figure}
\begin{figure}[htb] 
\centering 
\includegraphics[height=5cm,width=8cm]{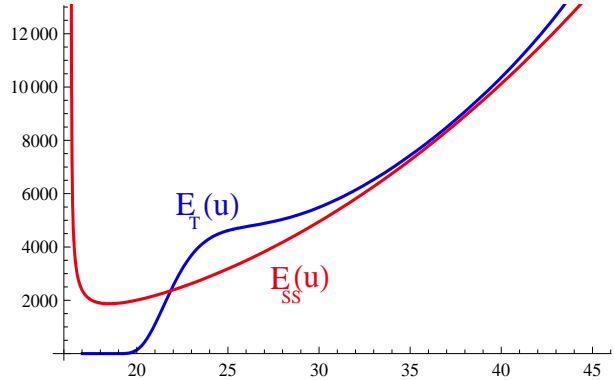}
\caption{The energy density integrands $E_{\rm SS}(u)$ and $E_{\rm T}(u)$.  
The rapid rise of the latter in the IR is clearly seen. The divergence 
between the two curves in the asymptotic region, 
$u \gtrsim u_{\rm max}$, is due to a nonzero $T_-$.}
\label{fig.10}
\end{figure}
In \fig{fig.9} we have plotted numerical solutions for $h(u)$ and
$T(u)$ for $u_0=17$ \footnote{Similar behaviour is seen for values of
$u_0 \gtrsim 14$. Below $u_0 \sim 14$, however, the energy difference
becomes very small and even reverses sign. This may be connected with
the breakdown of the approximate action in this region, similar to the
observation of a zero quark mass at $u_0 \sim 13.01$.}. For comparison
with the SS model, we have also plotted $h_{\rm SS}$ after adjusting
the value of $u_0$ for it to produce the same value of the asymptotic
brane-antibrane separation. The required value turns out to be
$u_0=16.4$. The corresponding energy density integrands, $E_{\rm
T}(u)$ and $E_{\rm SS}(u)$, have been plotted in
\fig{fig.10}. We can clearly see the rapid rise of $E(u)$ in the
IR, the subsequent flattening out and finally the power-law rise in
the asymptotic region. Using $u_{\rm max}=35.32$ \footnote{This is the
value at which $T(u)$ vanishes. The asymptotic form, \eq{sol-t}, fits
the numerically computed $T(u)$ in the range $33 \leq u \leq u_{\rm
max}$ to better than a percent with the parameter values
$h_0=0.179,~T_+=28904,~T_-=-0.0937$.}, numerical evaluation of the
integrals gives $(E_{\rm T}-E_{\rm SS})=-300.3$. Therefore, the
solution with the tachyon taken into account corresponds to a lower
energy state.

\section{\label{condensate}The chiral condensate}

By the standard dictionary of AdS/CFT \cite{AGMOO,EW1,BKL,BKLT}, once
we have identified $T_-$ with the quark mass parameter, we should
identify $T_+$ with the chiral condensate. However, it is not clear
that the standard rules apply to the present case of a boundary theory
which is not a CFT and has a scale. Moreover, the fact that there is
no known lift of \db~to $11$-dimensions forces an essential cut-off in
the theory with flavours. In fact, for a non-zero value of $T_-$, the
real cut-off is much lower, as we have seen from numerical
computations in the last section. Despite these difficulties, we will
assume that the identification of sources in the boundary theory with
boundary values of bulk fields holds in the theory with cut-off.

There is an additional difficulty in the present case. As we have seen
above, the desired solutions have only one independent parameter,
which we take to be $T_-$. The other three parameters, $T_+,~h_0$ and
$h_1$ should then be considered to be functions of $T_-$. Thus, the
chiral condensate cannot be computed naively by varying the on-shell
flavour brane action w.r.t. $T_-$, since this would also include
contributions from the variation of the other three parameters with
$T_-$. The one-parameter solutions that we have found constitute the
most general class of space-time independent solutions with the
specified boundary conditions \footnote{These boundary conditions are
(i) vanishing tachyon and fixed brane-antibrane separation asymptotically
and (ii) divergent tachyon and vanishing brane-antibrane separation at
some point in the bulk.}. Therefore, if we only want to make a
variation of $T_-$ only, we must go out of the present one-parameter
class of solutions to more general solutions, which are space-time
dependent, in addition to being dependent on $u$, and have enough
parameters. These solutions to $(u,x)$-dependent equations should have
the same singularities at $u=u_0$ as the solutions in
\eq{sol-la} and \eq{sol-ta}. Moreover, the asymptotic solutions should
have the form of \eq{sol-t} and \eq{sol-l} with $x$-dependent
coefficients. If solutions satisfying these conditions exist and have
enough parameters, then we can make the required variation of $T_-$
only and identify $T_+$ as the condensate in a coherent state formed
from fluctuations of $T$ and $h$ (scalar mesons) around the ground
state with broken chiral symmetry. Specializing to the $x$-independent
case, after varying the on-shell action, then, gives us the condensate
in the vacuum state. What we, therefore, need to do is to analyse the
$x$-dependent case to see if the required solutions exist. This is
what we will do next.

\subsection{\label{xdependence}Action for $(u,x)$-dependent $T$ and $h$}

The full $(u,x)$-dependent action for tachyon and brane-antibrane
separation is given by
\be
S=-\frac{2V_4}{R^9}\int d^4x \int du \ u^{13/4} \ 
V(T) \sqrt{d_T} \ \sqrt{{\rm det}(1+K)}, 
\label{xuaction}
\ee
where $K$ is the matrix with the
elements
\bea
K^{\mu}_{~\nu} &=& \frac{f}{4Q}\partial^{\mu}h\partial_{\nu}h
+\frac{u^{-3/2}}{Q}\partial^{\mu}T\partial_{\nu}T, \nonumber \\
K^{\mu}_{~u} &=& \frac{f}{4Q}h'\partial^{\mu}h
+\frac{u^{-3/2}}{Q}T'\partial^{\mu}T, \nonumber \\
K^u_{~\mu} &=& \frac{fu^{3/2}}{4d_T}h'\partial_{\mu}h
+\frac{1}{d_T}T'\partial_{\mu}T, \nonumber \\
K^u_{~u} &=& 0.
\label{kmatrix}
\eea
To look for a generalization of the $x$-independent solutions for
equations of motion derived from this action, the most obvious thing
to do is to generalize the earlier solutions by making all parameters
functions of $x$. In particular, this means making $u_0$, the place
where the flavour brane and antibrane meet, a function of $x$. For $u
\sim u_0$, expansion of this solution around a constant $u_0$ is
singular, since it involves arbitrary higher powers of
$1/(u-u_0)$. Therefore, we do not expect analysis of \eq{xuaction} by
expanding in small fluctuations around the $x$-independent solution to
work for $u$ close to $u_0$. This is confirmed by explicit fluctuation
calculations in Appendix \ref{fluctuations}. We need to go beyond
small fluctuations analysis of \eq{xuaction} and this requires us to
get an exact expression for the determinant in terms of space-time
derivatives of $T$ and $h$.

A direct calculation of ${\rm det}(1+K)$ is tedious, but the
calculation can be simplified using a trick which has been described
in Appendix \ref{det}, where a rather simple expression for the
determinant has been obtained. The complete $5$-dimensional action
then reads
\be
S=-\frac{2V_4}{R^9}\int d^4x \int du \ u^{13/4} \ 
V(T) \sqrt{\Delta_T}, 
\label{xuaction1}
\ee
where $\Delta_T=d_T\Delta$ and we have defined 
\be
\Delta \equiv 1+\beta_1(\partial T)^2+\beta_2(\partial h)^2+
2\beta_3(\partial h.\partial T)
+\beta_4[(\partial T)^2(\partial h)^2-(\partial h.\partial T)^2].
\label{delta}
\ee
The $\beta$'s are given by
\be
\beta_1=\frac{u^{-3/2}}{Q}(1-\frac{T^{\prime 2}}{d_T}), \ \
\beta_2=\frac{f}{4Q}(1-\frac{f u^{3/2} h^{\prime 2}}{4d_T}), \ \
\beta_3=-\frac{f h' T'}{4Qd_T}, \ \
\beta_4=\beta_1\beta_2-\beta_3^2.
\label{beetas}
\ee
As a check on the action \eq{xuaction1}, we note that it reduces to
the action \eq{new-t1} if $T$ and $h$ are $x$-independent. Also, it
correctly reproduces the action \eq{s} which only retains terms that
are quadratic in space-time derivatives of $T$ and $h$. This latter
action was derived independently by expanding det$(1+K)$ in powers of
$K$ and retaining only the first nontrivial correction.

The equations of motion that follow from the action \eq{xuaction1} are
rather complicated and have been derived in Appendix \ref{det},
\eq{xut} and \eq{xuh}. As we did in the $x$-independent case, we will
solve these equations in the two limiting cases of large $u$ and $u
\sim u_0$.

\medskip

\noindent{\large\underline{$u \rightarrow u_{\rm max}$:}} In this limit, 
$h(u,x)$ goes to a fixed value $h_0(x)$, which is assumed to be a
slowly varying function of $x$. We will also assume that $T$ and all
its derivatives are small in this limit. Then the equations \eq{xut}
and \eq{xuh} can be approximated as
\bea
-\left(u^4~T'(u,x)\right)'+(h_0(x))^2~u^4~T(u,x) &=& 0, \\
\left(u^{\frac{11}{2}}h'(u,x)\right)' &=& 0.
\eea
The space-time derivatives are comparatively suppressed by powers of
$1/u$ and hence have been ignored. These equations are identical to
\eq{asym-t} and \eq{asym-l} and so have solutions similar to
\eq{sol-t} and \eq{sol-l}, but now with parameters that are functions
of $x$:
\bea
T(u,x) &=& \frac{1}{u^2}(T_+(x) e^{-h_0(x) u}+T_-(x) e^{h_0(x) u}), \nn
h(u,x) &=& h_0(x)-h_1(x) u^{-9/2}.
\label{uvsol}
\eea

\medskip

\noindent{\large\underline{$u \rightarrow u_0$:}} The analysis in this
limit is somewhat more involved. We assume an ansatz similar to the
solutions \eq{sol-la} and \eq{sol-ta}, but now with $x$-dependent
$u_0$ and coefficients:
\bea 
h(u,x) &=& \rho_0(x)(u-u_0(x))^{1/2}+\rho_1(x) (u-u_0(x))^{3/2}+\cdots, \nn
T(u,x) &=& \sigma_0(x) (u-u_0(x))^{-2}+\sigma_1(x)(u-u_0(x))^{-1}+\cdots.
\label{ansatz}
\eea
As consequence of this ansatz, one can show that 
\bea
\del_\mu h &=& -h'[\del_\mu u_0-\frac{2\del_\mu \rho_0}{\rho_0}(u-u_0)
+\cdots], 
\label{hfromansatz} \\
\del_\mu T &=& -T'[\del_\mu u_0+\frac{\del_\mu \sigma_0}{2\sigma_0}(u-u_0)
+\cdots].
\label{tfromansatz}
\eea
These relations are correct to the order shown. Putting all this in
the equation of motion for $T$, \eq{xut}, we see that this equation is
satisfied to the leading order provided the following condition holds:
\be
\frac{13}{4u_0}-\frac{\sqrt{\pi}}{2}\sigma_0 \rho_0^2
=u_0^{-3/2}\del_\mu (u_0^{-3/2}\del^\mu u_0)-
\frac{1}{2}u_0^{-3}\del^\mu u_0 \frac{\del_\mu (u_0^{-3}(\del u_0)^2)}
{1+u_0^{-3}(\del u_0)^2}.
\label{tcond}
\ee
In obtaining this we have set $f_0=1$. Similarly, from \eq{xuh} one
gets the condition
\be
\sigma_0=\frac{\sqrt{\pi}}{4}(u_0^{3/2}+u_0^{-3/2}(\del u_0)^2).
\label{hcond}
\ee

If $u_0$ is a constant independent of $x$, then from equations
\eq{tcond} and \eq{hcond} one gets
\be
\sigma_0=\frac{\sqrt{\pi}}{4}u_0^{3/2}, \ \ 
\rho_0=\sqrt{\frac{26}{\pi u_0}}u_0^{-3/4}.
\label{xindepnorms}
\ee
These reproduce the $x$-independent solutions in \eq{sol-la} and
\eq{sol-ta}, remembering that we have set $f_0=1$. Let us now consider 
a small fluctuation around this constant solution. Linearizing the
equations \eq{tcond} and \eq{hcond} in fluctuations, we get
\be
\delta\sigma_0(x)=\frac{3\sqrt{\pi}}{8}u_0^{1/2}\delta u_0(x), \ \
\delta\rho_0(x)=-\frac{4u_0^{-13/4}}{\sqrt{26\pi}}(\del^2+\frac{65}{8}u_0)
\delta u_0.
\label{linear}
\ee
Now, clearly we could choose the fluctuation $\delta u_0(x)$ to be
such that $\delta\rho_0(x)$ vanishes. Under such an infinitesimal
change of $u_0$, $\sigma_0$ would change, but not $\rho_0$. It is this
kind of greater freedom in independently varying the parameters of the
solution that we have wanted. Presumably in higher orders the
situation gets better because there are more terms in the ansatz
\eq{ansatz} and for each coefficient there is some freedom because of 
the space-time dependence. It would be nice to analyse the higher
order terms, but that is beyond the scope of this work. Here we will
assume that the introduction of space-time dependence as above can
give us the required freedom to do the calculation of the condensate
as follows.

Finally, let us compare the solution \eq{xindepnorms}, \eq{linear}
with the solution obtained by the singular perturbation expansion in
Appendix \ref{fluctuations}, \eq{b16}. Expanding \eq{ansatz} around
constant $u_0$ solution to the lowest nontrivial order in $\epsilon
\equiv (u-u_0)$ and comparing with \eq{b13}, we get the relations
\bea
\varphi_0(x) &=& 2~\delta u_0(x), \hspace{1.1cm}
\varphi_1(x)=\frac{1}{\sigma_0}(\delta\sigma_0(x)+\sigma_1 \delta u_0(x)), \nn
\vartheta_0(x) &=& -\frac{1}{2}\delta u_0(x), \qquad
\vartheta_1(x)=\frac{1}{\rho_0}(\delta\rho_0(x)-
\frac{3}{2}\rho_1\delta u_0(x)).
\label{compare}
\eea
These relations involve not only the leading order parameters
\eq{xindepnorms} of the constant solution, but also the nonleading 
parameters $\sigma_1,~\rho_1$, which are given by
\be
\sigma_1=\frac{\sigma_0}{6u_0}, \qquad \rho_1=-\frac{5\rho_0}{8u_0}.  
\label{nonleading}
\ee
Using \eq{xindepnorms}-\eq{nonleading}, one can show that the
equations in \eq{b16} are satisfied. This equivalence is, however,
only formal. As we have argued above, the method given in this section
is the correct one to use since it does not involve a singular
expansion in arbitrarily high powers of $1/(u-u_0)$.
 
\subsection{Condensate in terms of the tachyon solution}

To derive an expression for the condensate, we calculate the variation
of the action in \eq{xuaction1} under a general variation of $T$ and
use the equation of motion \eq{xut} to reduce it to a boundary term:
\be
\delta S=-\frac{2V_4}{R^9}\int d^4x \frac{V(T)u^{13/4}}{\sqrt{d_T}}
T'(u,x)\delta T(u,x)|_{u=u_{\rm max}}.
\label{onshells1-1}
\ee
We have ignored terms with space-time derivatives because from now on
we will be specializing to the $x$-independent case, except in the
variation $\delta T$, so these terms will drop out.  Only the UV
boundary contributes to the on-shell action; there is no IR
contribution because the tachyon potential vanishes exponentially for
the diverging tachyon in the IR. We are only interested in retaining
the variation $\delta T_-(x)$, so we set $\delta T_+(x)$ to
zero. Doing this and using \eq{uvsol} in \eq{onshells1-1}, we get
the leading contribution for large $u_{\rm max}$,
\be
\delta S \approx \frac{2h_0V_4V(0)}{R^9}(T_+-T_- e^{2h_0u_{\rm max}})
\int d^4x~\delta T_-(x).
\label{onshells1-2}
\ee

On-shell brane actions have UV divergences which need to be removed by
the holographic renormalization procedure \footnote{For reviews, see
\cite{BFS,KS}.} to get finite answers for physical quantities. One adds
boundary counter terms to the brane action to remove the divergences,
following a procedure described in \cite{KBS}. Our on-shell action
\eq{onshells1-2} diverges as the cut-off is removed. This is because,
as discussed in section \ref{cutoff}, we are keeping
$T_+$ and $T_- e^{h_0 u_{\max}}$ fixed as the cut-off is removed and
the last term in \eq{onshells1-2} diverges as $e^{h_0 u_{\max}}$ in
this limit. The holographic renormalization procedure has been
developed for examples with CFT boundary theories. Since, with the
$D8$-branes present, there is no $11$-dimensional description
available to us, it is not clear that the procedure described in
\cite{KBS} is applicable to the present case. We will proceed on the
assumption that this is the case. Therefore, to subtract the UV
divergent term in \eq{onshells1-2}, we will add the following counter
term to the boundary action,
\be
S_{\rm ct}=\frac{V_4V(0)}{R^9}\int d^4x~\sqrt{-\gamma}~
h(u,x)T^2(u,x)|_{u_{\max}},
\label{ctaction}
\ee
where $\gamma=-u_{\max}^8$ is the determinant of the metric on the
$8$-dimensional boundary orthogonal to the slice at $u=u_{\max}$. Note
that the counter terms must be even in powers of the tachyon because
of gauge symmetry. Using the solution \eq{uvsol} and retaining only the
parameter $T_-(x)$, we find that the variation of the counter term
action is
\be
\delta S_{\rm ct}=\frac{2h_0V_4V(0)}{R^9}(T_++T_- e^{2h_0u_{\rm max}})
\int d^4x~\delta T_-(x).
\label{ctaction}
\ee
Adding to \eq{onshells1-2}, the divergent term drops out and we get
the variation of the renormalized action
\be
\delta S_{\rm renorm} \approx \frac{4h_0V_4V(0)}{R^9}T_+ 
\int d^4x~\delta T_-(x).
\label{renormaction}
\ee 
Note that the variation of the renormalized action is twice as large as
it would have been if we had simply dropped the divergent term
\footnote{In \eq{onshells1-2}, it is inconsistent to drop the term 
proportional to $T_-$ in the limit of large cut-off, holding $T_+$ and
$T_- e^{h_0 u_{\max}}$ fixed. In fact, it is the $T_-$ term that
dominates in the action \eq{onshells1-2} in this limit. Taking a
different limit that allows one to simply drop this term creates
difficulties in the calculation of the pion mass, see section
\ref{GOR-relation}. Consistency with the chiral condensate calculation
then demands that the term proportional to $T_+T_-$ be dropped in the
pion mass calculation since it is smaller than the $T_+^2$ term.} in
\eq{onshells1-2}.

\begin{figure}[htb] 
\centering 
\includegraphics[height=8cm,
width=14cm]{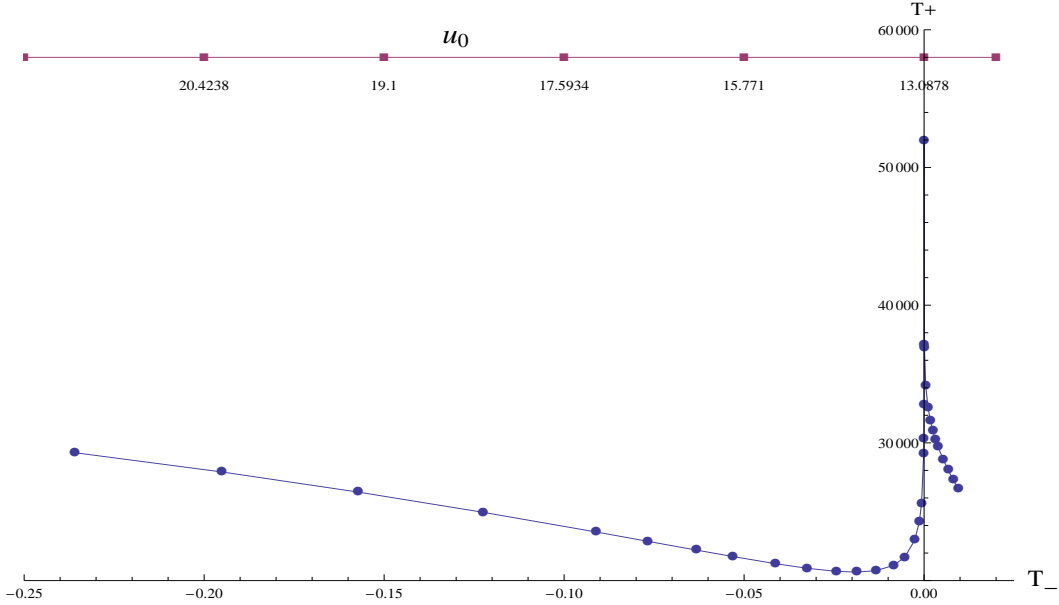}
\caption{$T_+$ as a function of $T_-$.}  
\label{fig.11}
\end{figure}

We are now ready to calculate an expression for the chiral condensate
in terms of the parameters of the tachyon solution. The parameters
$T_{\pm}$ are dimensionless. To construct a parameter of dimension
mass from $T_-$, we introduce a scale $\mu$ and define $m_q=\mu
|T_-|$. Then, identifying the chiral condensate $\chi \equiv <\bar{q_L} q_R>$, with $\delta S_{\rm renorm}/\mu \delta T_-(x)$, we get
\be
\chi \approx \frac{4h_0 V_4 V(0)}{\mu R^9}T_+
\label{cc}
\ee
We see that the parameter $T_+$ determines the
condensate. \fig{fig.11} shows a plot of $T_+$ as a function of $T_-$
for $T_- \sim 0$.
$T_+$ seems to attain a maximum value at $T_-=0$ and drops off
rapidly, at least for small values of $|T_-|$.

\section{\label{meson}The meson spectra}

In this section we will discuss the spectra for various low spin
mesons which are described by the fluctuations of the flavour branes
around the classical solution \footnote{For a general review of mesons
in gauge/gravity duals, see \cite{EEKT}.}. The action for the
fluctuations of the gauge fields can be computed from
\eq{with-t1}. Parametrizing the complex tachyon $\tau$ in terms of its
magnitude and phase, $\tau=Te^{i\theta}$, we get the following action,
correct to second order in the fluctuations:
\bea
\Delta S_{\rm gauge} &=& -\int d^4x~du \biggl[a(u) A_u^2 + b(u) A_\mu^2 + 
c(u)\left((F^V_{\mu\nu})^2+(F^A_{\mu\nu})^2\right)+e(u) F^A_{\mu u} A^\mu \nn
&& + d(u) \left((F^V_{\mu u})^2+(F^A_{\mu u})^2\right) \biggr],
\label{gaugefluc1} \\
a(u) &=& R^{-15} V_4 V(T)u^{13/4}
\frac{T^2}{\sqrt{d_T}},
\label{gaugefluc2} \\
b(u) &=& R^{-3} V_4 V(T)u^{7/4} \sqrt{d_T} \frac{T^2}{Q}
\left(1+\frac{f^2T^2h^2h'^2}{4d_T}u^3\right), 
\label{gaugefluc3} \\
c(u) &=& \frac{R^3}{8} V_4 V(T)u^{1/4} \sqrt{d_T},
\label{gaugefluc4} \\
d(u) &=& R^{-9} V_4 V(T)u^{7/4} \frac{Q}{4\sqrt{d_T}},
\label{gaugefluc5} \\
e(u) &=& R^{-6} V_4 V(T)u^{13/4} 
\frac{fT^2hh'}{2\sqrt{d_T}}.
\label{gaugefluc5}
\eea
Here $F^V_{\mu\nu}$ is the usual field strength for the vector gauge
field $V=(A_1+A_2)$ and $F^A_{\mu\nu}$ is the field strength for the
gauge-invariant combination of the axial vector field and the phase of
the tachyon, $A=(A_1-A_2-\del\theta)$. However,
\be
F^V_{\mu u}=-F^V_{u \mu}=\del_\mu V_u-R^3 \del_u V_\mu, \quad 
F^A_{\mu u}=-F^A_{u \mu}=\del_\mu A_u-R^3 \del_u A_\mu.
\label{mixed}
\ee
The relative factor of $R^3$ simply reflects the change of variables
\eq{redef}. 

The gauge field $V_\mu(x,u)$ gives rise to a tower of
vector mesons while the fields $A_\mu(x,u)$ and $A_u(x,u)$, which are
gauge invariant, give rise to towers of axial and pseudoscalar
mesons. Notice that the coefficients $a(u)$, $b(u)$ and $e(u)$ vanish
if the tachyon is set to zero. In the absence of the tachyon, the
vector and axial vector mesons acquire masses because of a nonzero
$d(u)$, but there is always a massless ``pion'' \footnote{Strictly
speaking, for the $U(1)$ case under discussion, this pseudoscalar is
the $\eta'$. It is massless here because of the $N_c \ra \infty$ limit
in which we are working.}. The presence of the tachyon is thus
essential to give a mass to the ``pion''. Also note that with the
tachyon present, the masses of the vector and axial vector mesons are
in principle different.

\subsection{Vector mesons}

We will be using the gauge $V_u=0$. Expanding in modes, we have
\be
V_\mu(x,u)=\sum_m V^{(m)}_\mu(x) W_m(u), 
\label{vmodes}
\ee
where $\{W_m(u)\}$ form a complete sets of basis functions. These
satisfy orthonormality conditions which will be determined presently.
The fields $\{V^{(m)}_\mu(x)\}$ form a tower of vector mesons in the
physical $(3+1)$-dimensional space-time. In terms of these fields, the
vector part of the action \eq{gaugefluc1} takes the form,
\be
\Delta S^V_{\rm gauge}=-\int d^4x~\sum_{m,n}\biggl[Q_{mn}^V
F^{V(m)}_{\mu\nu}F^{V(n)\mu\nu}+L_{mn}^V V^{(m)}_{\mu}V^{(n)\mu}\biggr], 
\label{vaction1}
\ee
where $F^{V(m)}_{\mu\nu}$ are the usual $(3+1)$-dimensional
$U(1)$-invariant field strengths for the vector potentials
$\{V^{(m)}_\mu\}$. Also, we have defined
\be
Q_{mn}^V=\int du~c(u) W_m(u)W_n(u), \quad
L_{mn}^V=R^6 \int du~d(u) W'_m(u)W'_n(u).
\label{vnorm}
\ee
In addition, we choose the basis functions $\{W_m(u)\}$ to satisfy the
eigenvalue equations
\be
-R^6\left(d(u) W'_m(u)\right)'=2 \lambda^V_m c(u) W_m, 
\label{veigeneqn}
\ee
Using these we see that
\be
L_{mn}^V=\frac{1}{2}\biggl[R^6 \biggl(d(u) W'_m(u)W_n(u)\biggr)_{\del u}+
2 \lambda^V_m Q_{mn}^V]+ m \leftrightarrow n, 
\label{veigen1}
\ee
where, as in the previous section, $\del u$ refers to boundaries in
the $u$-direction.

Note that a potential zero mode in the vector sector \footnote{A zero
mode is defined as a mode which has zero eigenvalue and goes to a
constant at infinity.} can be gauged away using the residual symmetry
of making $u$-independent gauge transformations, which is still
available after fixing the gauge $V_u=0$. This is because a zero mode
in this sector can only have a single scalar degree of freedom. This
follows from the requirement of finiteness of the the action,
\eq{vaction1}, which cannot be satisfied since the coefficient of the
field strength term blows up for a zero mode. Hence its field strength
must vanish, leaving behind only a longitudinal degree of freedom.

For the nonzero modes we may, without loss of generality, choose
\be
Q_{mn}^V = \frac{1}{4} \delta_{mn}, 
\label{vnorm}
\ee
which, on using \eq{veigen1}, gives
\be
L_{mn}^V=\frac{1}{2} \lambda^V_m \delta_{mn}.
\label{veigen2}
\ee
Using \eq{vnorm} and \eq{veigen2} in \eq{vaction1}, we get 
\be
\Delta S^V_{\rm gauge}=-\int d^4x~\sum_m \biggl[\frac{1}{4}
F^{V(m)}_{\mu\nu}F^{V(m)\mu\nu}+ \frac{1}{2} \lambda^V_m
V^{(m)}_{\mu}V^{(m)\mu}\biggr].
\label{vaction2}
\ee

\subsection{Axial vector and pseudoscalar mesons}

As we have already noted, $A_\mu$ and $A_u$ are gauge
invariant. Expanding in modes, we have
\be 
A_{\mu}(x,u)=\sum_m A^{(m)}_{\mu}(x) P_m(u),\quad
A_u(x,u)=\sum_m\phi^{(m)}(x)S_m(u),
\label{amodes}
\ee
where $\{P_m(u)\}$ and $\{S_m(u)\}$ form complete sets of basis
functions. These satisfy orthonormality conditions which will be
determined presently. The fields $\{A_{\mu}^{(m)}(x)\}$ and
$\{\phi^{(m)}(x)\}$ form towers of axial vector and pseudoscalar mesons
in the physical $(3+1)$-dimensional space-time. In terms of these
fields, the axial-vector and pseudoscalar part of the action
\eq{gaugefluc1} takes the form,
\bea
\Delta S^A_{\rm gauge} &=& -\int d^4x~\sum_{m,n}\biggl[
\frac{1}{2} \delta_{mn} \lambda^\phi_m \phi^{(m)}\phi^{(n)} 
+Q_{mn}^A F^{A(m)}_{\mu\nu}F^{A(n)\mu\nu} +L_{mn}^A
A^{(m)}_{\mu}A^{(n)\mu} \nn 
&& +K_{mn} \partial_{\mu}\phi^{(m)}\partial^{\mu}\phi^{(n)}+J_{mn} A^{(m)\mu}
\partial_{\mu}\phi^{(n)}\biggr],
\label{aaction1}
\eea
where $F^{A(m)}_{\mu\nu}$ are the usual $(3+1)$-dimensional
$U(1)$-invariant field strengths for the axial vector potentials
$\{A_{\mu}^{(m)}\}$. Also, we have defined
\bea
Q_{mn}^A &=& \int du~c(u) P_m(u)P_n(u), \nn
L_{mn}^A &=& \int du~\biggl(R^6 d(u) P'_m(u)P'_n(u)
+(b(u)+\frac{1}{2}R^3e'(u))P_m(u) P_n(u)\biggr), \nn
J_{mn} &=& \int du~\biggl(e(u)P_m(u)-2R^3d(u)P_m^{\prime}(u)\biggr)S_n(u),\nn
K_{mn} &=& \int du~d(u)S_m(u)S_n(u),
\label{anorm}
\eea
and used the orthonormality condition in the pseudoscalar sector
\bea
\int du~a(u) S_m(u)S_n(u) = \frac{1}{2} \lambda^\phi_m \delta_{mn}.
\label{pionmass}
\eea
In addition, we choose the basis functions $\{P_m(u)\}$ to satisfy the
eigenvalue equations
\be
-R^6\biggl(d(u)P'_m(u)\biggr)'+\biggl(b(u)+\frac{1}{2}R^3e'(u)\biggr)P_m(u)
=2 \lambda^A_m c(u) P_m(u).
\label{aeigeneqn}
\ee
Using these we see that
\be
L_{mn}^A = \frac{1}{2}\biggl[R^6\biggl(d(u) P'_m(u)P_n(u)\biggr)_{\del u}+
2 \lambda^A_m Q_{mn}^A\biggr]+ m \leftrightarrow n, 
\label{aeigen1}
\ee
where, as before, $\del u$ refers to boundaries in the $u$-direction.

We note that because of the last term in \eq{aaction1}, the
longitudinal component of $A^{(m)}_\mu$ and $\phi^{(m)}$ mix. So we
need to define new field variables in terms of which the action
\eq{aaction1} is diagonal. Before we do that, let us first note that
the axial vector potential $A_\mu(x, u)$ has a possible zero mode
provided the corresponding $(3+1)$-dimensional field strength
vanishes, for reasons explained in the previous subsection. Hence the
zero mode, which we shall denote by $A^{(0)}_\mu$, can only have a
longitudinal component. The zero mode is gauge-invariant and, because
of its mixing with the pseudoscalars, plays a special role. Let us see
this in some detail.

The zero mode $A^{(0)}_\mu$ is conjugate to the eigenfunction $P_0(u)$
which satisfies the equation
\be
-R^6\biggl(d(u)P'_0(u)\biggr)'+\biggl(b(u)+\frac{1}{2}R^3e'(u)\biggr)P_0(u)=0.
\label{zmode}
\ee
If there is no solution to this equation, then the zero mode does not
exist and we should proceed directly to diagonalize the action
\eq{aaction1}. If, however, a solution $P_0(u)$ to this equation
exists and is such that it goes to a constant at infinity, then the
zero mode $A^{(0)}_\mu$ exists. Since it is purely longitudinal, for a
reason identical to that discussed in the vector case, we make this
explicit by writing it in terms of a pseudoscalar field,
$A^{(0)}_\mu=\del_\mu
\phi^{(0)}(x)$. The terms in the action
\eq{aaction1} which contain $\phi^{(0)}(x)$ can be separated
out. These terms are: $$L_{00}^A
\del_\mu \phi^{(0)} \del^\mu \phi^{(0)} +\sum_m J_{0m} \del_\mu
\phi^{(m)} \del^\mu \phi^{(0)}.$$ The sums over the indices $m,n$ no
longer include the zero mode. Also, we have used $L_{m0}^A=L_{0m}^A=0$
for $m \neq 0$, which follows from \eq{aeigen1} using the fact
that $\lambda_0^A=0$ and the boundary terms vanish because $P_m(u)$
vanishes sufficiently fast at infinity. Without loss of generality, we
may choose $L_{00}^A=1/2$ (to get the normalization of the kinetic
term of $\phi^{(0)}$ right). Then, we can rewrite the above as 
\be
\frac{1}{2} \del_\mu \pi \del^\mu \pi-\frac{1}{2} \sum_{m,n} J_{0m}J_{0n}
\del_\mu \phi^{(m)} \del^\mu \phi^{(n)},
\label{pion}
\ee
where $\pi \equiv (\phi^{(0)}+\sum_m J_{0m} \phi^{(m)})$. 

With the zero modes explicitly separated out in this way, for the
nonzero modes we may, without loss of generality, choose
\be
Q_{mn}^A=\frac{1}{4} \delta_{mn}, 
\label{anorm}
\ee
which, on using \eq{aeigen1}, gives
\be
L_{mn}^A=\frac{1}{2} \lambda^A_m \delta_{mn}.
\label{aeigen2}
\ee
Putting \eq{pion}, \eq{anorm} and \eq{aeigen2} in the action \eq{aaction1}, 
we get
\bea
\Delta S^A_{\rm gauge} &=& -\int d^4x~\biggl[\sum_m \biggl(
\frac{1}{2} \lambda^\phi_m \phi^{(m)}\phi^{(m)} 
+\frac{1}{4} F^{A(m)}_{\mu\nu}F^{A(m)\mu\nu} 
+ \frac{1}{2} \lambda^A_m A^{(m)}_{\mu}A^{(m)\mu}\biggr)  \nn
&& +\frac{1}{2} \del_\mu \pi \del^\mu \pi
+\sum_{m,n} \biggl(\tilde K_{mn}\partial_{\mu}\phi^{(m)} \partial^{\mu}\phi^{(n)} 
+J_{mn} A^{(m)\mu} \partial_{\mu}\phi^{(n)}\biggr)\biggr]
\label{aaction2},
\eea
where $\tilde K_{mn}=(K_{mn}-\frac{1}{2} J_{0m}J_{0n})$. The above
action describes a massless particle, $\pi$, besides other massive
particles. The existence of this massless particle depends on the
existence of a solution to the equation (\ref{zmode}), satisfying the
normalization condition 
\be
R^6\biggr(d(u) P_0(u)P_0'(u)\biggr)_{\del u}=\frac{1}{2}.
\label{pnorm}
\ee
Later we will see that the existence of the desired solution $P_0(u)$
depends on the absence of a non-normalizable part in $T(u)$.

To diagonalize the action \eq{aaction2} for the massive modes, we
define the new variables
\be
A^{(m)}_\mu=\tilde A^{(m)}_\mu-\sum_n (\lambda_m^A)^{-1} J_{mn} 
\partial_{\mu}\phi^{(n)}.
\label{unmix}
\ee 
Putting in \eq{aaction2}, we get
\bea
\Delta S^A_{\rm gauge} &=& -\int d^4x~\biggl[\sum_m \biggl(
\frac{1}{2} \lambda^\phi_m \phi^{(m)}\phi^{(m)} 
+\frac{1}{4} F^{A(m)}_{\mu\nu}F^{A(m)\mu\nu} 
+ \frac{1}{2} \lambda^A_m \tilde A^{(m)}_{\mu} \tilde A^{(m)\mu}\biggr) \nn
&& +\frac{1}{2} \del_\mu \pi \del^\mu \pi 
+\sum_{m,n} K'_{mn} \partial_{\mu}\phi^{(m)} 
\partial^{\mu}\phi^{(n)} \biggr], \nn
\label{aaction3}
\eea
where $K'_{mn}=(\tilde K_{mn}-\frac{1}{2} \sum_p (\lambda_p^A)^{-1}
J_{pm} J_{pn})$. The modes have now been decoupled. To get the
standard action for massive pseudoscalars we may, without loss of
generality, set
\be
K'_{mn}=\frac{1}{2} \delta_{mn}=K_{mn}-\frac{1}{2} J_{0m}
J_{0n}-\frac{1}{2} \sum_p (\lambda_p^A)^{-1} J_{pm} J_{pn}
\label{kay}
\ee
This condition can be rewritten in a more conventional form as
follows. We define
\be
\psi_m(u) \equiv \sum_n (\lambda_n^A)^{-1} P_n(u) J_{nm}+P_0(u) J_{0m},
\label{psi}
\ee
and using (\ref{aeigeneqn}) note that it satisfies the equation
\be 
-R^6\biggl(d(u)\psi'_m(u)\biggr)'+\biggl(b(u)+
\frac{1}{2}R^3e'(u)\biggr) \psi_m(u)
=\frac{1}{2} e(u)S_m(u)+R^3\biggl(d(u)S_m(u)\biggr)'.
\label{psieq}
\ee
Using (\ref{psi}) in (\ref{kay}), we get 
\be
\delta_{mn}=\int du \biggl(d(u)S_m(u)(S_n(u)+R^3\psi'_n(u))
-\frac{1}{2}e(u)S_m(u)\psi_n(u)\biggr)+m \leftrightarrow n.
\label{psi2}
\ee
In terms of new variables defined by 
\be
S_m(u) \equiv R^3\eta'_m(u), \quad \quad 
\theta_m(u) \equiv \psi_m(u)+\eta_m(u),
\label{newvar}
\ee
\eq{psi2} can be written as
\be
\int du~\eta'_m(u) \biggl(R^6d(u) \theta'_n(u)-\frac{1}{2}
R^3e(u) (\theta_n(u)-\eta_n(u)) \biggr)+m \leftrightarrow n=\delta_{mn}.
\label{theta2}
\ee
Moreover, in terms of these variables the differential equation
\eq{psieq} reads
\be
 -R^6\biggl(d(u)\theta'_m(u)\biggr)'+\biggl(b(u)+\frac{1}{2}R^3e'(u)\biggr)
\biggl(\theta_m(u)-\eta_m(u)\biggr)-\frac{1}{2} R^3e(u) \eta'_m(u)=0,
\label{theta1}
\ee
From these two equations one can obtain the orthonormality condition
\bea
&& \int du \biggl(R^6d(u) \theta'_m(u) \theta'_n(u) + 
(b(u)+\frac{1}{2}R^3e'(u))(\theta_m(u)-\eta_m(u))(\theta_n(u)-\eta_n(u)) \nn
&& -\frac{1}{2} R^3e(u) \eta'_m(u)(\theta_n(u)-\eta_n(u))
-\frac{1}{2} R^3e(u) \eta'_n(u)(\theta_m(u)-\eta_m(u)) \biggr)
=\frac{1}{2} \delta_{mn}. \nn
\label{pseudo1}
\eea
Also, rewriting (\ref{pionmass}) in terms of the new variables, we
have 
\be
R^6\int du~a(u) \eta'_m(u) \eta'_n(u)=\frac{1}{2} \lambda_m^\phi \delta_{mn}.
\label{pseudo2}
\ee
Finally, \eq{theta2} and \eq{pseudo2} give
\be
R^6a(u) \eta'_n(u)=\lambda_n^\phi \biggl(R^6d(u) \theta'_n(u)-
\frac{1}{2} R^3e(u) (\theta_n(u)-\eta_n(u)) \biggr).
\label{pseudo3}
\ee
Equations (\ref{theta1}) and (\ref{pseudo3}) are the final form of the
eigenvalue equations and (\ref{pseudo1}) and (\ref{pseudo2}) are the
orthonormality conditions in the pseudoscalar sector.

It is interesting to note from (\ref{theta1}) that if $\eta$ is
constant, then the variable $(\theta-\eta)$ satisfies a differential
equation that is identical to the equation (\ref{zmode}) satisfied by
the zero mode $P_0$. Also, using (\ref{theta1}) and (\ref{pseudo1})
one can show that for constant $\eta$, $(\theta-\eta)$ satisfies the
normalization condition (\ref{pnorm}). From (\ref{pseudo3}) it follows
that if $\eta$ is constant, the eigenvalue $\lambda^\phi$
vanishes. Thus, the presence of a massless pseudoscalar can be
naturally considered to be identical to the question of the existence
of a solution to the equations (\ref{theta1})-(\ref{pseudo3}) with
zero eigenvalue, and so it becomes a part of the spectrum in the
pseudoscalar tower of states. Hence, the action in this sector can be
written in the form
\bea
\Delta S^A_{\rm gauge} &=& -\int d^4x~\sum_m \biggl[ 
\frac{1}{4} F^{A(m)}_{\mu\nu}F^{A(m)\mu\nu} 
+ \frac{1}{2} \lambda^A_m \tilde A^{(m)}_{\mu} \tilde A^{(m)\mu} \nn
&& \hspace{3cm}+\frac{1}{2}\partial_{\mu}\phi^{(m)}\partial^{\mu}\phi^{(m)}
+\frac{1}{2} \lambda^\phi_m \phi^{(m)}\phi^{(m)} \biggr].
\label{aaction3}
\eea
Note that we have dropped the field $\pi(x)$, but extended the sum
over $m$ to cover a possible zero mode as well. If there is a solution
to the equations \eq{theta1}-\eq{pseudo3} with constant $\eta_0$ and
$\lambda^\phi_0=0$, then a massless pion field will reappear as the
zero mode $\phi^{(0)}$ in the pseudoscalar tower. Otherwise, the
lowest mode in this sector will be massive, whose mass can be computed
as in the following subsection.

\subsection{\label{GOR-relation}Relation between pion mass and 
non-normalizable part of tachyon}

In this subsection we will derive a relation between the pion mass and
the non-normalizable part of tachyon parametrized by $T_-$. This will
give us further evidence for identifying the parameters $T_+$ and
$T_-$ with the chiral condensate and quark mass respectively. We first
note that for $T(u)=0$, $a(u)$ vanishes and hence $\lambda^\phi_m$
also vanishes by \eq{pseudo3}. However, as we will see from the
following calculations, $T(u)=0$ is a sufficient condition, but not
necessary to guarantee the presence of a massless pion. The necessary
condition is that the non-normalizable piece in $T(u)$ should be
absent, i.e. $T_-=0$.

Let us assume that $T(u) \neq 0$ so that $a(u) \neq 0$. Then,
\eq{pseudo3} can be used to solve for $\eta_m(u)$ in terms of
$\psi_m(u)$, which is related to $\theta_m(u)$ and $\eta_m(u)$ by
\eq{newvar}. We get,
\be
\eta'_m(u)=\frac{\lambda^\phi_m}{a(u)-\lambda^\phi_m d(u)}
\biggl(d(u)\psi'_m(u)-\frac{e(u)}{2R^3}\psi_m(u)\biggr)
\label{etam}
\ee
Let us now denote by $\lambda^\phi_0$ the lowest mass eigenvalue. The
corresponding eigenfunctions are $\psi_0(u)$ and $\eta_0(u)$. Assuming
$\lambda^\phi_0 \ll a(u)/d(u)$ \footnote{This approximation can be
justified a posteriori by the solution because the eigenvalue
$\lambda^\phi_0$ turns out to be parametrically much smaller by a
factor of $1/R^3$, see \eq{mineigen2}, compared to the ratio
$a(u)/d(u)$.}, we can approximate the above equation for $\eta_0(u)$:
\be
\eta'_0(u) \approx \frac{\lambda^\phi_0}{a(u)}
\biggl(d(u)\psi'_0(u)-\frac{e(u)}{2R^3}\psi_0(u)\biggr)
\label{eta0}
\ee
If we know $\psi_0(u)$, then using the above in \eq{pseudo2} we can
compute the mass. Now, $\psi_0(u)$ satisfies the following
differential equation, which can be obtained from \eq{theta1} using
\eq{eta0} and the approximation $\lambda^\phi_0 \ll a(u)/d(u)$:
\be
-R^6\biggl(d(u)\psi'_0(u)\biggr)'+\biggl(b(u)+
\frac{1}{2}R^3e'(u)\biggr)\psi_0(u) \approx 0.
\label{psi01}
\ee
Also, using \eq{psi01} and the approximation under which it was
obtained, the normalization condition on $\psi_0(u)$ given by
\eq{pseudo1} can be approximated as
\be
R^6d(u)\psi'_0(u)\psi_0(u)|_{u=u_{\rm max}} \approx \frac{1}{2}.
\label{psinorm}
\ee

These equations cannot be solved analytically in general. However,
analytic solutions can be obtained in the IR and UV regimes. In the UV
regime, for $u \lesssim u_{\rm max}$, we use \eq{sol-t} and \eq{sol-l}
to approximate the coefficients in \eq{psi01}; we get
\be
b(u) \approx \frac{V_4 V(0)}{R^3}uT^2(u), \quad 
d(u) \approx \frac{V_4 V(0)}{4R^9}u^{5/2}, \quad 
e(u) \approx \frac{9 V_4 V(0)}{4R^6}h_0h_1u^{-3/2}T^2(u). 
\label{uv}
\ee
In writing these, we have used $f(u) \approx 1$, which is a good
approximation for large $u$.  We see that we can clearly neglect
$e(u)$ compared to $b(u)$ in \eq{psi01}, while $b(u)$ is itself
negligible compared to $d(u)$. Using these approximations in
\eq{psi01} and \eq{psinorm} then gives
\be
-\biggl(u^{\frac{5}{2}}\psi'_0(u)\biggr)' \approx 0, \quad
\frac{V_4V(0)}{4R^3} u^{\frac{5}{2}}\psi'_0(u)\psi_0(u)|_{u=u_{\rm max}} 
\approx \frac{1}{2},
\label{psi02}
\ee
which are solved by 
\be
\psi_0(u) \approx c_0-\frac{1}{3c_0}\frac{4R^3}{V_4V(0)}u^{-3/2}.
\label{psiuv}
\ee
Here $c_0$ is a parameter which is related to the pion decay constant.
This can be argued by analysing the $4$-d axial current correlator and
using AdS/CFT along the lines of \cite{EKSS,RP}. Using the AdS/CFT
dictionary, one can compute the axial current correlator from the
action \eq{aaction1}, evaluated on-shell, by differentiating twice
with respect to the transverse part of the axial vector field on the
UV boundary. This is the source which couples to the axial current on
the boundary. The source arises from the same zero mode solution,
$P_0(u)$, which we discussed in connection with a possible zero mode
(the pion) in the longitudinal component of the axial gauge
field. $P_0(u)$ satisfies the equation \eq{zmode}, which is identical
to that satisfied by $\psi_0(u)$, \eq{psi01}. However, the boundary
condition now is different; it is the boundary condition for a source,
$P_0(u_{\max})=1$. In addition, one imposes the condition
\be
R^6 d(u) P'_0(u)P_0(u)|_{u=u_{\rm max}} \approx \frac{f_\pi^2}{2},
\label{sourcenorm}
\ee
which is required to reproduce the correct zero momentum axial current
correlator \cite{EKSS,RP}. This follows from the action
\eq{aaction1}. Now, $P_0(u)$ satisfies \eq{zmode} and the condition
\eq{sourcenorm} if we set $P_0(u)=f_\pi \psi_0(u)$. 
Then, requiring $P_0(u_{\rm max})=1$ gives $c_0=1/f_\pi$.

In the IR regime, $u \gtrsim u_0$, we use \eq{sol-la} and \eq{sol-ta}
to approximate the coefficients in \eq{psi01}; we get 
\be
b(u) \approx \frac{\pi^{3/2} V_4 u_0^{17/4}}{26 R^3}\frac{V(T)}{(u-u_0)^4},
\quad d(u) \approx \frac{13 V_4 u_0^{9/4}}{32 \sqrt{\pi} R^9}V(T), \quad
e(u) \approx \frac{13 V_4 u_0^{9/4}}{16 \sqrt{\pi} R^6}\frac{V(T)}{(u-u_0)}.
\label{ir}
\ee
In writing these, we have used $f(u_0) \approx 1$, which is a good
approximation for large $u_0$.  Using $dV(T)/du=T'(u)V'(T)$, we see
that $b(u)$ and $R^3e'(u)$ both go as $(u-u_0)^{-4}$ in this
regime. However, the coefficient of the latter is suppressed by a
relative factor of $u_0^{-1/2}$, so for large $u_0$ we may neglect it
compared to $b(u)$. But, unlike in the UV regime, $b(u)$ cannot be
neglected compared to $d(u)$. In fact, this term is crucial for getting
a nontrivial solution. In this regime, then, the leading terms in
equation \eq{psi01} give
\be
\psi'_0(u) \approx \frac{32\pi R^6 u_0^{1/2}}{169}\frac{\psi_0(u)}{(u-u_0)},
\label{psi02}
\ee
which has the solution
\be
\psi_0(u) \approx \tilde{c_0} (u-u_0)^{\frac{32\pi R^6 u_0^{1/2}}{169}},
\label{psiir}
\ee
where $\tilde{c_0}$ is an integration constant. Note that the
normalization condition remains unchanged and cannot be used here
because it receives contribution only from the UV end due to the
exponentially vanishing tachyon potential for large $T(u)$ at the IR
end.

Let us now consider the formula, \eq{pseudo2}, for the lowest mode, using
which one can compute the eigenvalue $\lambda^\phi_0$:
\be
R^6\int^{u_{\rm max}}_{u_0} du~ a(u) (\eta'_0(u))^2=\frac{1}{2} \lambda_0^\phi.
\label{lowestmass}
\ee
Using $a(u) \approx \frac{\sqrt{\pi} V_4 u_0^{19/4}}{8 R^{15}}
\frac{V(T)}{(u-u_0)}$ in the IR and \eq{psiir} in \eq{eta0}, we
see that $\eta'_0(u) \propto \psi_0(u)$ vanishes very rapidly as $u
\rightarrow u_0$, with a power which grows as $u_0^{1/2}$ for large
$u_0$. Moreover, since $V(T)$ vanishes exponentially for large $T$,
the IR region makes a negligible contribution to the
integral. Therefore, it is reasonable to calculate the integral by
substituting the UV estimate of the integrand in it. In the UV region,
$a(u) \approx \frac{V_4 V(0)}{R^{15}} u^4T^2(u)$. Moreover, in this
region the second term on the right hand side of \eq{eta0} can be
neglected. So, we get
\bea
\frac{1}{2} \lambda_0^\phi=
R^6\int^{u_{\rm max}}_{u_0} du~a(u) (\eta'_0(u))^2 & \approx & 
R^6(\lambda^\phi_0)^2\int^{u_{\rm max}}_{\tilde{u}_0} du~\frac{d^2(u)}{a(u)} 
(\psi'_0(u))^2 \nonumber \\
& \approx & (\lambda^\phi_0)^2 \kappa
\int^{u_{\rm max}}_{\tilde{u}_0} \frac{h_0~du}{(T_+ e^{-h_0u}+
T_- e^{h_0u})^2}, 
\nonumber
\eea
where $\tilde{u}_0 > u_0$ avoids the IR region in the integral and
we have defined 
\be
\kappa \equiv \frac{f_\pi^2 R^9}{4 h_0 V_4 V(0)}.
\label{kappa}
\ee 
The integral is easily done, giving
\be
\lambda^\phi_0 \approx \frac{1}{\kappa}\frac{(T_+ e^{-h_0\tilde{u}_0}+
T_- e^{h_0\tilde{u}_0})(T_+ e^{-h_0u_{\rm max}}+
T_- e^{h_0u_{\rm max}})}{e^{h_0(u_{\rm max}-\tilde{u}_0)}
-e^{-h_0(u_{\rm max}-\tilde{u}_0)}}.
\label{mineigen0}
\ee
From our numerical solutions we see that it is possible to choose
$\tilde{u}_0$ to be relatively large and also satisfy the
conditions $|T_+| e^{-h_0\tilde{u}_0} \gg |T_-| e^{h_0\tilde{u}_0}$
and $e^{h_0(u_{\rm max}-\tilde{u}_0)} \gg e^{-h_0(u_{\rm
max}-\tilde{u}_0)}$. For such values of the parameters, then, to a
good approximation \eq{mineigen0} gives
\be
\lambda^\phi_0 \approx \frac{1}{\kappa}(T_+T_-+T_+^2 e^{-2h_0u_{\rm max}}).
\label{mineigen1}
\ee
Now, let us tune $u_{\max}$ to large values. We will do this in a
manner consistent with the inequality \eq{umax}. As explained in
section \ref{cutoff}, one way of maintaining this inequality is to
keep $|T_+|$ and $|T_-|e^{h_0u_{\max}}$ fixed as $u_{\max}$ becomes
large. In that case, the second term on the right hand side of
\eq{mineigen1} becomes exponentially smaller than the first term as
the cut-off is increased beyond some value. We may then neglect this
term compared with the first term. This gives 
\be
\lambda^\phi_0 \approx \frac{1}{\kappa}T_+T_-.
\label{mineigen2}
\ee
Finally, using $\lambda^\phi_0=m_\pi^2$ and \eq{cc} in this relation,
we get
\be
m_\pi^2 \approx \frac{m_q\chi}{f_\pi^2},
\label{gor}
\ee
This is the well-known Gell-Mann$-$Oakes$-$Renner formula, up to a
factor of $2$.

\section{\label{discussion}Summary and Discussion}

This paper further explores our proposal \cite{DN} of a modified SS
model, which includes the degree of freedom associated with the open
string tachyon between the flavour branes and antibranes. Here we have
extended the analytic treatment of various aspects of the problem and
supplemented it with extensive numerical calculations. We have argued
that taking the tachyon into account is essential for the consistency
of the setup and shown numerically that the solution which includes
the tachyon is energetically favoured. Our modification preserves the
nice geometric picture of chiral symmetry breaking of the SS model and
at the same time relates chiral symmetry breaking to tachyon
condensation; the tachyon becomes infinitely large in the infrared
region where the joining of the flavour branes signals chiral symmetry
breaking.

We have identified a parameter in the non-normalizable part of the
tachyon field profile with the quark mass. It is important to stress
that this is the only tunable parameter in the modified SS model. It
can be traded for the asymptotic brane-antibrane separation or the
location of the point in the bulk where the brane and antibrane
join. This provides a natural explanation for the latter parameter,
which is also present in the SS model, but in that model it doesn't
find any counterpart in the QCD-like theory at the boundary. In this
paper we have presented numerical evidence to show that the point
where the brane and antibrane meet is monotonically shifted towards
ultraviolet as we tune the mass parameter to larger values. It would
seem, therefore, that in our model a brane-antibrane pair disappears
from the bulk consistently with a quark flavour becoming infinitely
massive.

The presence of a non-normalizable part in the tachyon solution
requires us to introduce an ultraviolet cut-off. The cut-off is needed
not only because this part grows as one moves towards the ultraviolet
region, as in any standard AdS/CFT example that includes a
non-normalizable solution, but also because the asymptotic form of
the solution is derived from an approximate equation which is valid
only for small values of the tachyon. Therefore, the asymptotic
solution itself is not valid beyond a certain maximum value of the
holographic coordinate. We have presented sufficient numerical
evidence of this phenomenon. Removing the ultraviolet cut-off, then,
requires tuning the mass parameter to zero. We have explained one
scheme by which this can be done. This scheme gives an exponential
dependence on the cut-off to the mass parameter, similar to that
discussed recently in \cite{HHLY}. The quark mass arises from an
apparently very different mechanism in this work and the cut-off is
related to the location of a $D6$-brane that is present in this model.
It would be interesting to see if there is any connection between this
model and our model.

Once we have identified the quark mass as a parameter in the
non-normalizable part of the tachyon, it is natural to expect, by the
usual AdS/CFT rules, the normalizable part of the tachyon solution to
give rise to the chiral condensate. To derive an expression for it,
however, we need to go beyond the space-time independent solutions of
section 2. As we have seen, this requires an exact expression for the
$5$-dimensional action for tachyon and brane-antibrane separation
fields which are now taken to depend on space-time as well as the
holographic coordinate. We have derived this action in this
paper. Using the generalized solutions to the equations for this
action, then, one can compute the chiral condensate. However, one also
needs to add counter terms to the boundary brane action to remove from
it contributions that diverge when the cut-off is removed.

We have studied in detail the fluctuations of flavour gauge fields on
the brane-antibrane system. These give rise to vector, axial vector
and pseudoscalar towers of mesons, which become massive through a kind
of higgs mechanism, except for the pions. These arise from a
gauge-invariant combination of the tachyon phase and the longitudinal
zero mode of the axial vector field. We have shown that the pions
remain massless, unless a quark mass (non-normalizable part of the
tachyon solution) is switched on. For small quark mass, we have
derived an expression for the mass of the lowest pseudoscalar meson in
terms of the chiral condensate and shown that it satisfies the
Gell-Mann$-$Oakes$-$Renner relation. The vector and axial vector
spectra are expected to be non-degenerate because they arise from
eigenvalue equations with different tachyon contributions. We have not
computed these spectra, but it would be interesting to see whether
they have the Regge behaviour for large masses.

A non-zero quark mass is essential to correctly reproduce
phenomenology in the low-energy sector of QCD. Therefore, our modified
SS model can be the starting point of a more quantitative version of the
phenomenology initiated in \cite{SS1}. For this purpose, our treatment
needs to be extended to the non-abelian case, which should be a
straightforward exercise. The correct tachyon brane-antibrane action
for curved directions transverse to the branes is not known. It is
important to have such an action since this would extend the
applicability of the present treatment to such interesting cases as
e.g. the antipodal configuration of the flavour brane system and its
connection with massless quarks. Another direction in which the
present ideas can be extended is to discuss this model at finite
temperature and describe the chiral symmetry restoration transition
and study the phase diagram in some detail. The connection of chiral
symmetry breaking with tachyon condensation seems fascinating and a
deeper understanding would be useful. Finally, baryons have been
discussed in the SS model. It turns out that they have a very small
size. This may change in the presence of the tachyon. This is because
in the presence of the tachyon, the flavour energy momentum tensor is
concentrated away from the infrared region where the branes meet. In
other words, there is a new scale provided by the quark mass. It would
be very interesting to investigate whether this effect makes any
difference to the baryon size.

\gap5

\noindent{\bf Acknowledgment}

\gap3

It is a pleasure to thank Gautam Mandal and Spenta Wadia for
discussions.

\gap5

\appendix

\section{\label{A}Overlapping $D8$-$\overline{D8}$-brane system}

In this case the appropriate DBI action is
\bea
S &=& -\int d^9\sigma~g_s V(T)~e^{-\phi}
\left(\sqrt{-\rm{det}~A_L}+\sqrt{-\rm{det}~A_R}~\right), \nn
(A_{i})_{ab} &=& g_{MN} \del_ax^M_i \del_bx^N_i+F^i_{ab}+ 
\frac{1}{2}\biggl((D_a\tau (D_b\tau)^*+(D_a\tau)^* D_b\tau)\biggr),
\label{olping1}
\eea
where $D_a\tau=\del_a \tau-i(A_{L,a}-A_{R,a})\tau$. The classical
equation for the profile of the magnitude $T$ of the tachyon $\tau$
can be obtained from (\ref{eq-t}) by substituting $h=0$ in it
everywhere. We get
\be
\left(\frac{u^{\frac{13}{4}}}{\sqrt{d_T}} T'(u)\right)' 
= \frac{u^{\frac{7}{4}}f(u)^{-1}}{\sqrt{d_T}}
\frac{V'(T)}{V(T)},
\label{olping2}
\ee
where now $d_T=f(u)^{-1} u^{-3/2}+{T'(u)}^2$. In the UV region,
assuming $T$ is small for large $u$, we can approximate this equation
as
\be
\biggl(u^4 T'(u)\biggl)'=-\pi u^{5/2} T(u),
\label{olping3}
\ee
where we have used the universal small $T$ expansion, $V(T)={\cal T}_8
(1-\frac{\pi}{2}T^2+\cdots)$. The general solution
\footnote{Equation (\ref{olping3}) can be solved exactly in terms of 
the Bessel functions $H^{(1)}$ and $H^{(2)}$. Here we give only the
leading term.} to this equation is
\be
T(u)=u^{-13/8}\biggl(c_1 {\rm cos}(4\sqrt{\pi}u^{1/4})+ c_2 {\rm
sin}(4\sqrt{\pi}u^{1/4})\biggr)+\cdots,
\label{olping4}
\ee
where $c_1$ and $c_2$ are arbitrary constants. Both the independent
solutions in this case are normalizable, so the interpretation of one
of the parameters corresponding to a source for the quark mass term is
not clear. In view of this, it is not clear how to apply the general
treatment of \cite{CKP} to this case. 

In the IR region, a singular tachyon solution is obtained only for $u
\sim u_k$. In this region $f(u)^{-1}$ blows up as $(u-u_k)^{-1}$ and 
this drives a singularity in the tachyon. Both the potentials in
(\ref{pot1}) and (\ref{pot2}) exhibit singular solutions, although the
solutions and the nature of singularity are different. For the potential 
(\ref{pot1}) we find the solution
\be
T(u)=\biggl(\pi+\frac{39}{2\sqrt{u_k}}\biggr)^{-1/2}
{\rm ln}\frac{1}{(u-u_k)}+b_1+\cdots,
\label{olping5}
\ee 
while for (\ref{pot2}) we get
\be
T(u)=b_2(u-u_k)^{-\alpha}+\cdots,
\label{olping6}
\ee
where $b_1$ and $b_2$ are arbitrary constants and
$\alpha=\frac{4\pi\sqrt{u_k}}{39}$. As in the case with nonzero
brane-antibrane separation, the IR solution for which the tachyon
blows up exhibits a smaller number of independent parameters than the
UV solution, one in the IR as opposed to two in the UV in the present
case.  A solution with two independent parameters in the IR exists
(for any potential), but this solution is finite:
\be
T(u)=T_0+T_1 (u-u_k)^{1/2}+(\frac{2}{3\sqrt{u_k}} 
+\frac{T_1^2}{2}) \frac{V'(T_0)}{V(T_0)}(u-u_k)+\cdots.
\label{olping7}
\ee
Here $T_0$ and $T_1$ are the two arbitrary parameters.

\section{\label{fluctuations}Scalar fluctuations}

Here we will assume that $T(u,x)$ and $h(u,x)$ are weakly dependent on
$x^\mu$ and expand ${\rm det}(1+K)$ in \eq{xuaction} in powers of
space-time derivatives. The action correct to quadratic terms in the
derivatives is
\bea
S &=& -\frac{2V_4}{R^9}\int d^4x \int du \ u^{13/4} \ 
V(T) \sqrt{d_T} \biggl[1+\frac{u^{-3/2}}{2Q} \biggl\{
(1-\frac{T^{\prime 2}}{d_T})(\partial T)^2 \nonumber \\
&& +(1-\frac{\frac{1}{4} f u^{3/2} h^{\prime 2}}{d_T})
\frac{1}{4} f u^{3/2}(\partial h)^2-
\frac{f u^{3/2} h' T'}{2 d_T} (\partial h) . (\partial T) 
\biggr\} \biggr],
\label{s}
\eea
where $d_T$ is given by \eq{new-t2}, with $T(u)$ replaced by $T(u,x)$
and $h(u)$ by $h(u,x)$. Also, the notation $(\partial T)^2$ stands for
$\eta^{\mu\nu} \partial_{\mu}T(u,x) \partial_{\nu}T(u,x)$; similar
expressions hold for $(\partial h)^2$ and $(\partial h) . (\partial
T)$. For the expansion in derivatives to be valid, we
must require the following conditions to be satisfied: (i) For large
values of $u$, near the cut-off $u_{\rm max}$, we must have $|\partial
T| << u_{\rm max}^{3/4}$ and $|\partial h| << 1$; (ii) For $u \sim
u_0$, we must have $|\partial T| << |T'| \sim (u-u_0)^{-3}$ and
$|\partial h| << |hT| \sim (u-u_0)^{-3/2}$.

Let us now consider small fluctuations around the $x$-independent
solutions. We write $T(u,x)=T_c(u)+T_q(u,x)$ and
$h(u,x)=h_c(u)+h_q(u,x)$, where $T_c(u)$ and $h_c(u)$ are the
$x$-independent solutions of the classical equations \eq{eq-t},
\eq{eq-l}. We now expand the above action and retain only terms up to
second order in the fluctuations $T_q(u,x)$ and $h_q(u,x)$. We get
\bea
S &=& -\frac{2V_4}{R^9} \int d^4x \int_{u_0}^\infty du \ A\sqrt{d_c}\biggl[1+
\biggl\{\frac{V_c'}{V_c}T_q
+\frac{1}{d_c}(\frac{1}{4} fu^{3/2}h_c'h_q'+T_c'T_q'+h_cT_c^2h_q+h_c^2T_cT_q)
\biggr\} \nonumber \\
&& ~~~~~~~~+\biggl\{\frac{V_c''}{2V_c}T_q^2
+\frac{V_c'}{V_c d_c}(\frac{1}{4} fu^{3/2}h_c'h_q'T_q+T_c'T_q'T_q
+h_cT_c^2h_qT_q+h_c^2T_cT_q^2) \nonumber \\
&& ~~~~~~~~~~~~+\frac{1}{2 d_c}(\frac{1}{4} fu^{3/2}h_q^{\prime 2}
+T_q^{\prime 2}+T_c^2h_q^2+h_c^2T_q^2+4 h_cT_ch_qT_q) 
-\frac{1}{2d_c^2}(\frac{1}{4} fu^{3/2}h_c'h_q'
\nonumber \\ && ~~~~~~~~~~~~+T_c'T_q'+h_cT_c^2h_q+h_c^2T_cT_q)^2 
+\frac{u^{-3/2}}{2Q_0} \biggl((1-\frac{T_c^{\prime 2}}{d_c})(\partial T_q)^2 
\nonumber \\
&& ~~~~~~~~~~~~-\frac{f u^{3/2}}{2 d_c}h_c'T_c' (\partial h_q) . (\partial T_q) 
+(1-\frac{\frac{1}{4} f u^{3/2} h_c^{\prime 2}}{d_c})
\frac{1}{4} f u^{3/2}(\partial h_q)^2 \biggr) \biggr\}+\cdots \biggr],
\label{b3}
\eea
where we have used the notation $V_c=V(T_c)$, $d_c=d_{T_c}$, and
$A=u^{13/4} \ V_c$.  As before, a prime denotes derivative w.r.t.
$u$, except on $V_c$, for which it denotes a derivative w.r.t.  its
argument. The part of this action linear in fluctuations, $S_1$, which
arises from the term in the first curly brackets above, is given by
\be
S_1=-\frac{2V_4}{R^9}\int d^4x \int_{u_0}^\infty du \ A \biggl[\frac{V_c'}{V_c}
\sqrt{d_c}T_q+\frac{1}{\sqrt{d_c}}(\frac{1}{4} f u^{3/2} h_c'h_q'+T_c'T_q'
+h_c T_c^2 h_q+h_c^2 T_c T_q)\biggr],
\label{s1}
\ee
It is easy to verify that $S_1$ leads to the background equations
\eq{eq-t} and \eq{eq-l}. This part of the action, therefore, vanishes,
except for a boundary term. It is this boundary term that gives rise
to the chiral condensate.

The term in the second curly brackets becomes $S_2$, the action
quadratic in fluctuations, after some manipulations. First, we open
the square in the coefficient of $1/2 d_c^2$ term and combine it with
the term just before it. That is, we have,
\bea
&& \frac{1}{2 d_c}(\frac{1}{4} fu^{3/2}h_q^{\prime 2}
+T_q^{\prime 2}+T_c^2h_q^2+h_c^2T_q^2+4 h_cT_ch_qT_q) \nonumber \\
&& ~~~~~~~~~~~~~~~~~~~~~~~~~~~~~~~~-\frac{1}{2d_c^2}(\frac{1}{4} fu^{3/2}h_c'h_q'
+T_c'T_q'+h_cT_c^2h_q+h_c^2T_cT_q)^2 \nonumber \\
&& =\frac{1}{2 d_c} \biggl\{(1-\frac{\frac{1}{4} f u^{3/2} h_c^{\prime 2}}{d_c})
\frac{1}{4} f u^{3/2}h_q^{\prime 2}+(1-\frac{T_c^{\prime 2}}{d_c})T_q^{\prime 2}
+(1-\frac{h_c^2 T_c^2}{d_c})(h_c^2 T_q^2+T_c^2 h_q^2) \nonumber \\
&& ~~~~~~~~+2(2-\frac{h_c^2 T_c^2}{d_c})h_cT_ch_q T_q\biggr\} 
 -\frac{1}{d_c^2}\biggl\{\frac{1}{4} f u^{3/2}h_c'(T_c'h_q'T_q'
+T_c^2h_ch_qh_q'+h_c^2T_ch_q'T_q) \nonumber \\
&& ~~~~~~~~+T_c'h_cT_c(T_cT_q'h_q+h_cT_q'T_q)\biggr\} 
\label{b4}
\eea
Furthermore, we can rewrite
\bea
A\frac{V_c'}{V_c\sqrt{d_c}}T_c' T_q T_q' && \sim 
V_c'\left(\frac{u^{13/4}T_c'}{\sqrt{d_c}}\right)\left(\frac{T_q^2}{2}\right)'
\nonumber \\
&& \rightarrow -A\sqrt{d_c}\left[\frac{V_c''}{V_c}\frac{T_c^{\prime 2}}{d_c}
+\frac{V_c'}{V_c}\left(\frac{h_c^2T_c}{d_c}+\frac{V_c'}{V_c}
(1-\frac{T_c^{\prime 2}}{d_c})\right)\right]\frac{T_q^2}{2},
\label{b5}
\eea 
where in the last step we have done an integration by parts over $u$,
used the equation of motion \eq{eq-t} for $T_c$, $h_c$ and ignored a
possible boundary term since it is quadratic in fluctuations and so
will not contribute to the calculation of the condensate. A similar
manipulation gives 
\bea
-\frac{A}{d_c\sqrt{d_c}}T_c'h_c^2T_cT_qT_q' && \sim  
-V_c\left(\frac{u^{13/4}T_c'}{\sqrt{d_c}}\right)
\left(\frac{h_c^2T_c}{d_c}\right)\left(\frac{T_q^2}{2}\right)' \nonumber \\
&& \rightarrow A\sqrt{d_c}\left[\left(\frac{V_c'}{V_c}+
\frac{h_c^2T_c}{d_c}\right)
\frac{h_c^2T_c}{d_c}+\frac{T_c'}{d_c}\left(\frac{h_c^2T_c}{d_c}\right)'
\right]\frac{T_q^2}{2}.
\label{b6}
\eea
Combining the above with the other three $T_q^2/2$ terms, we find its
net coefficient to be
\be
A\left\{\left(\frac{V_c''}{V_c}- (\frac{V_c'}{V_c})^2 \right)
\left(1-\frac{T_c^{\prime 2}}{d_c}\right)\sqrt{d_c}+2\frac{V_c'}{V_c}
\frac{h_c^2T_c}{\sqrt{d_c}}+\frac{h_c^2}{\sqrt{d_c}}+
\frac{T_c'}{\sqrt{d_c}}\left(\frac{h_c^2T_c}{d_c}\right)'\right\}
\label{b7}
\ee
Similarly, a partial integration using the equation of motion
\eq{eq-l} allows us to combine the two $h_q^2/2$ terms, giving its net
coefficient to be
\be
A\left\{\left(\frac{h_cT_c^2}{d_c}\right)'
\frac{\frac{1}{4}f u^{3/2} h_c'}{\sqrt{d_c}}+\frac{T_c^2}{\sqrt{d_c}}\right\}
\label{b8}
\ee
Collecting all this together, we get the action quadratic in fluctuations:
\bea
S_2 &=& -\frac{2V_4}{R^9}\int d^4x \int_{u_0}^\infty du \ A\biggl[
\frac{1}{2}c_1T _q^2+\frac{1}{2}c_2 h_q^2+\frac{1}{2}c_3 h_q^{\prime 2}
+\frac{1}{2}c_4 T_q^{\prime 2}
+c_5 h_q T_q+c_6 h_q'T_q' \nonumber \\
&& +c_7 h_q'T_q+c_8h_qT_q'+\frac{c_9}{8u^3Q_c}(\partial T_q)^2
+\frac{c_{10}}{4u^3Q_c}(\partial h_q) . (\partial T_q)
+\frac{c_{11}}{8u^3Q_c}(\partial h_q)^2 \biggr],
\label{b9}
\eea
where the coefficients $\{c_i\}$ are given by   
\bea
c_1 &=& \left(\frac{V_c'}{V_c}\right)'
\left(1-\frac{T_c^{\prime 2}}{d_c}\right)\sqrt{d_c}+2\frac{V_c'}{V_c}
\frac{h_c^2T_c}{\sqrt{d_c}}+\frac{h_c^2}{\sqrt{d_c}}+
\frac{T_c'}{\sqrt{d_c}}\left(\frac{h_c^2T_c}{d_c}\right)' \\
c_2 &=& \left(\frac{h_cT_c^2}{d_c}\right)'
\frac{\frac{1}{4}f u^{3/2} h_c'}{\sqrt{d_c}}
+\frac{T_c^2}{\sqrt{d_c}}, \\
c_3 &=& \frac{1}{\sqrt{d_c}}\biggl(1-\frac{\frac{1}{4} f u^{3/2} h_c^{\prime 2}}
{d_c}\biggr)\frac{1}{4}f u^{3/2}, \\
c_4 &=& \frac{1}{\sqrt{d_c}}\biggl(1-\frac{T_c^{\prime 2}}{d_c}\biggr), \\
c_5 &=& \frac{V_c'}{V_c}\frac{h_cT_c^2}{\sqrt{d_c}} 
+\biggl(2-\frac{h_c^2 T_c^2}{d_c}\biggr)\frac{h_cT_c}{\sqrt{d_c}}, \\
c_6 &=& -\frac{T_c'}{d_c\sqrt{d_c}}\frac{1}{4} f u^{3/2}h_c', \\
c_7 &=& \frac{1}{\sqrt{d_c}}\biggl(\frac{V_c'}{V_c}-\frac{h_c^2 T_c}{d_c}\biggr)
\frac{1}{4} f u^{3/2}h_c', \\
c_8 &=& -\frac{h_cT_c^2T_c'}{d_c\sqrt{d_c}}, \\
c_9 &=& 4u^{3/2}\sqrt{d_c} \biggl(1-\frac{T_c^{\prime 2}}
{d_c}\biggr), \\
c_{10} &=& -u^3\frac{f}{\sqrt{d_c}} h_c'T_c', \\
c_{11} &=& u^3f\sqrt{d_c}\biggl(1-\frac{\frac{1}{4} f u^{3/2} 
h_c^{\prime 2}}{d_c}\biggr), 
\eea
with $Q_c=(1+f u^{3/2} h_c^2 T_c^2)$. For later convenience, we have
explicitly written out a factor of $1/4u^3Q_c$ in the coefficients in
the last three terms in \eq{b9}.

This action mixes $T_q$ and $h_q$ and the equations of motion derived
from it reflect this mixing. After some manipulations, the equations
can be cast in the form
\bea
\partial^2 T_q &=& a_1 T_q+a_2 T_q'+a_3 T_q''+a_4 h_q+a_5 h_q', 
\label{b10} \\
\partial^2 h_q &=& b_1 h_q+b_2 h_q'+b_3 h_q''+b_4 T_q+b_5 T_q',
\label{b11}
\eea
where the coefficients $\{a_i\}$ and $\{b_i\}$are given by
\bea
a_1 &=& c_{10}(\bar{c_7}-c_5)+c_{11}c_1, \hspace{0.77cm}
a_2=c_{10}(\bar{c_6}+c_7-c_8)-c_{11}\bar{c_4}, \quad
a_3=c_{10}c_6-c_{11}c_4, \nn
a_4 &=& -c_{10}c_2+c_{11}(c_5-\bar{c_8}), \quad
a_5=c_{10}\bar{c_3}-c_{11}(\bar{c_6}-c_7+c_8), 
\label{b12-a}
\eea
and
\bea
b_1 &=& c_{10}(\bar{c_8}-c_5)+c_9c_2, \hspace{0.73cm}
b_2=c_{10}(\bar{c_6}-c_7+c_8)-c_9\bar{c_3}, \quad
b_3=c_{10}c_6-c_9c_3, \nn
b_4 &=& -c_{10}c_1+c_9(c_5-\bar{c_7}), \quad
b_5=c_{10}\bar{c_4}-c_9(\bar{c_6}+c_7-c_8).
\label{b12-b}
\eea
Here we have used the notation $\bar{c_i}=(Ac_i)'/A$. As usual, a
prime denotes a derivative with respect to $u$. Moreover,
$\partial^2=(-\partial_t^2+\partial_{\vec{x}}^2)$ is the flat
space-time laplacian. A possible term proportional to $h_q''$ is not
present in \eq{b10} because its coefficient, $(c_{10}c_3-c_{11}c_6)$, 
vanishes. Similarly, in \eq{b11} the term proportional to $T_q''$ is 
absent because its coefficient, $(c_{10}c_4-c_9c_6)$, vanishes.

The equations of motion derived from \eq{b9} are quite complicated
in general, but they simplify in the two asymptotic regimes of $u$.

\medskip

\noindent{\large\underline{$u \rightarrow u_{\rm max}$:}} In this limit, 
many of the $c_i$ are small because they have at least one factor of
$T_c$ or its derivatives in them. The exceptions are $c_1 \sim
h_0^2 u^{3/4},~c_3 \sim u^{9/4}/4,~c_4 \sim u^{3/4},~c_9 \sim
4 u^{3/4}$ and $c_{11} \sim u^{9/4}$. Retaining only the dominant
terms in the equations, we get
\bea
-\left(u^4~T_q'(u,x)\right)'+h_0^2~u^4~T_q(u,x) &=& 0, \\
\left(u^{\frac{11}{2}}h_q'(u,x)\right)' &=& 0.
\eea
The term involving space-time laplacian on the fluctuations can be
consistently neglected at the leading order since it is non-leading in
powers of $u$, as can be verified a posteriori. These equations are
identical to \eq{asym-t} and \eq{asym-l} and so have solutions similar
to \eq{sol-t} and \eq{sol-l}, but now with parameters that are
functions of $x$:
\bea
T_q(u,x) &=& \frac{1}{u^2}(T_{q+}(x) e^{-h_0 u}+T_{q-}(x) e^{h_0 u}), \nn
h_q(u,x) &=& h_{q0}(x)-h_{q1}(x) u^{-9/2}.
\label{sol-x}
\eea

\medskip

\noindent{\large\underline{$u \rightarrow u_0$:}} This limit is 
more involved, requiring a more detailed analysis. One expands $T_q$
and $h_q$ in powers of $\epsilon \equiv (u-u_0)$ with arbitrary
$x$-dependent coefficients.
\bea
T_q(u,x) &=& \frac{\sqrt{\pi}}{4} u_0^{3/2}
\epsilon^\omega \biggl(\varphi_0(x)+\epsilon~\varphi_1(x)+\cdots \biggr), \nn
h_q(u,x) &=& \sqrt{\frac{26}{\pi u_0}} u_0^{-3/4}
\epsilon^\tau \biggl(\vartheta_0(x)+\epsilon~\vartheta_1(x)+
\cdots \biggr),
\label{b13}
\eea
Here, and in the following, we have set $f_0=1$. One also needs to
expand the $a_i$'s and $b_i$'s in powers of $\epsilon$. Retaining
up to the first nonleading power in $\epsilon$, we get
\bea
a_1 &=& 8\xi \epsilon^{-1}(1+\frac{23\epsilon}{12u_0}), \ \
a_2=2\xi(1+\frac{2\epsilon}{u_0}), \ \
a_3=\frac{4 u_0^{-3/2}}{\pi}\xi \epsilon^3(1+\frac{23\epsilon}{12u_0}), \nn
a_4 &=& \frac{2\pi u_0^{11/4}}{\sqrt{26}}\xi \epsilon^{-7/2}
(1+\frac{65\epsilon}{24u_0}), \ \
a_5=\frac{4 u_0^{5/4}}{\sqrt{26}}\xi \epsilon^{-1/2}
(1+\frac{21\epsilon}{8u_0}),
\label{b14}
\eea
and 
\bea
b_1 &=& -3\xi \epsilon^{-1}(1+\frac{3\epsilon}{4u_0}), \ \
b_2=2\xi(1+\frac{2\epsilon}{u_0}), \ \
b_3=\frac{4 u_0^{-3/2}}{\pi}\xi \epsilon^3(1+\frac{23\epsilon}{12u_0}), \nn
b_4 &=& \frac{16\sqrt{26} u_0^{-11/4}}{\pi}\xi \epsilon^{3/2}
(-1+\frac{\epsilon}{24u_0}), \ \
b_5=-\frac{4\sqrt{26} u_0^{-11/4}}{\pi}\xi \epsilon^{5/2}
(1+\frac{\epsilon}{24u_0}), \nn
\label{b15}
\eea
where $\xi=-13 u_0^2/8$. Substituting these expansions in the
equations \eq{b10}, \eq{b11} and comparing different orders of
$\epsilon$, we see that a consistent solution exists only for
$\omega=-3$ and $\tau=-1/2$, and then we get
\be
\vartheta_0(x)=-\frac{1}{4}\varphi_0(x), \ \
\varphi_1(x)=\frac{5}{6u_0}\varphi_0(x), \ \
\vartheta_1(x)=\frac{1}{8\xi}(\del^2+\frac{65u_0}{32})\varphi_0(x).
\label{b16}
\ee
The first of these relations is precisely what is needed to think of
the leading terms in \eq{b13} as coming from expanding
$(u-u_0(x))^{-1}$ around a constant $u_0$. The last relation shows
that when $x$-dependence is allowed, not all coefficients get uniquely
determined. In fact, the part of $\varphi_0(x)$ annihilated by the
operator on the right hand side does not show up in $\vartheta_1(x)$.

The above analysis shows that perturbation expansion in ``small''
fluctuations around a constant $u_0$ is singular, although we have
obtained a solution by a formal expansion.

\section{\label{det}Calculation of the exact $(u,x)$-dependent action}

This involves calculating the determinant of the matrix $(1+K)$, whose
elements are given in \eq{kmatrix}. We will simplify this calculation
by making use of the following trick. Consider the family of
determinants, $D(\lambda) \equiv {\rm det}(1+\lambda K)$, where
$\lambda$ is an arbitrary parameter. We actually only need to
calculate $D(1)$, but this calculation can be reduced
essentially to the calculation of the inverse of the matrix
$(1+\lambda K)$, which turns out to be much easier than a direct
computation of the determinant. Consider the following:
\be
\frac{d}{d\lambda}D(\lambda)=D(\lambda)
{\rm tr}[(1+\lambda K)^{-1}K].
\ee
We can obtain $\Delta$ by integrating this equation, using the
boundary condition $D(0)=1$:
\be
{\rm ln}D(1)=\int_0^1 d\lambda~{D(\lambda)}^{-1}
\frac{d}{d\lambda}D(\lambda)=\int_0^1 d\lambda~
{\rm tr}[(1+\lambda K)^{-1}K]
\label{lnd}
\ee
This reduces the required calculation to finding the inverse matrix
$M(\lambda)=(1+\lambda K)^{-1}$, which may be done as follows. Using
the defining equation, $(1+\lambda K)M(\lambda)=1$, one can express
all components of $M$ in terms of ${M^\mu}_\nu$:
\be
{M^u}_\nu=-\lambda {K^u}_\mu {M^\mu}_\nu, \ \ {M^u}_u=1-\lambda^2 {K^u}_\mu
{K^\nu}_u{M^\mu}_\nu, \ \ {M^\mu}_u=-\lambda {M^\mu}_\nu {K^\nu}_u.
\label{mrelations}
\ee
Moreover, one can show that ${M^\mu}_\nu$ satisfies
\be
{P^\mu}_\sigma {M^\sigma}_\nu={{\delta}^\mu}_\nu, \ \ \
{P^\mu}_\sigma \equiv ({{\delta}^\mu}_\sigma+\lambda {K^\mu}_\sigma-
\lambda^2 {K^\mu}_u {K^u}_\sigma).
\label{inverse}
\ee
Thus, to find $M(\lambda)$ we need to find the inverse of the
${P^\mu}_\sigma(\lambda)$ matrix. First note that using \eq{kmatrix}
we can write
\be
{P^\mu}_\sigma(\lambda)={{\delta}^\mu}_\nu+\beta_1(\lambda) 
\del^\mu T \del_\nu T+
\beta_2(\lambda) \del^\mu h \del_\nu h+\beta_3(\lambda) 
(\del^\mu T \del_\nu h+\del^\mu h \del_\nu T),
\label{pmatrix}
\ee
where
\bea
\beta_1(\lambda) &=& \frac{\lambda u^{-3/2}}{Q}
(1-\lambda \frac{T^{\prime 2}}{d_T}), \qquad \qquad
\beta_2(\lambda)=\frac{\lambda f}{4Q}(1-\lambda \frac{f u^{3/2} 
h^{\prime 2}}{4d_T}), \nn
\beta_3(\lambda) &=& -\frac{\lambda^2 f h' T'}{4Qd_T}, \hspace{3cm}
\beta_4(\lambda)=\beta_1(\lambda)\beta_2(\lambda)-(\beta_3(\lambda))^2.
\label{lambeetas}
\eea
For $\lambda=1$ these reduce to the $\beta$'s in \eq{beetas}. Now,
from the general structure of the ${P^\mu}_\nu$ matrix, we can
parametrize the ${M^\mu}_\nu$ matrix as
\be
{M^\mu}_\nu(\lambda)={{\delta}^\mu}_\nu+\alpha_1(\lambda) 
\del^\mu T \del_\nu T+
\alpha_2(\lambda) \del^\mu h \del_\nu h+\alpha_3(\lambda) 
(\del^\mu T \del_\nu h+\del^\mu h \del_\nu T).
\label{mmatrix}
\ee
We have calculated the $\alpha$'s. They work out to be
\bea
\alpha_1(\lambda) &=& -\frac{1}{\Delta(\lambda)}[\beta_1(\lambda)+
\beta_4(\lambda)(\partial h)^2], \nn
\alpha_2(\lambda) &=& -\frac{1}{\Delta(\lambda)}[\beta_2(\lambda)+
\beta_4(\lambda)(\partial T)^2], \nn
\alpha_3(\lambda) &=& -\frac{1}{\Delta(\lambda)}[\beta_3(\lambda)-
\beta_4(\lambda)\partial h.\partial T].
\label{alpha}
\eea
Here $\Delta(\lambda)$ is a generalization of $\Delta$ defined in
\eq{delta}. It has the same form but with the above $\lambda$-dependent 
$\beta$'s replacing those in \eq{delta}. By definition,
$\Delta(1)=\Delta$.

Armed with the inverse matrix $M(\lambda)$, we can now compute the
trace on the right hand side of \eq{lnd}. Using \eq{mrelations}
and \eq{kmatrix}, we first note that
\be
{\rm tr}[(1+\lambda K)^{-1}K]={M^\mu}_\sigma(\lambda) 
\frac{d}{d\lambda}{P^\sigma}_\mu(\lambda).
\ee
Given the equations \eq{pmatrix}-\eq{alpha}, it is straightforward,
though tedious, to compute the right hand side of the above
equation. One gets the simple result
\be
{M^\mu}_\sigma(\lambda)
\frac{d}{d\lambda}{P^\sigma}_\mu(\lambda)={\Delta(\lambda)}^{-1}
\frac{d}{d\lambda}\Delta(\lambda).
\ee
It follows from this and \eq{lnd} that $D(1)=\Delta(1)=\Delta$. Hence
the complete $5$-dimensional action is that given in \eq{xuaction1}.

To compute the equations of motion for $T(u,x)$ and $h(u,x)$ that
follow from this action, we will need the following, which can be
easily calculated from the relation $\Delta_T=d_T\Delta$ and the
definition of $\Delta$ given in \eq{delta}:
\bea
\frac{1}{2} \frac{\del \Delta_T}{\del T'} &=& T'+\frac{fT'}{4Q}(\partial h)^2
-\frac{fh'}{4Q} \del T.\del h, \nn
\frac{1}{2} \frac{\del \Delta_T}{\del(\del_\mu T)} &=& d_T\beta_1 \del^\mu T
+d_T \beta_3\del^\mu h+\frac{u^{-3}}{4Q}\biggl(\del^\mu T(\partial h)^2-
\del^\mu h(\partial h.\partial T) \biggr), \nn
\frac{1}{2} \frac{\del \Delta_T}{\del T} &=& Th^2 \biggl[1-\frac{f^2u^{3/2}}
{4Q^2} \biggl(h^{\prime 2} (\del T)^2+T^{\prime 2} (\del h)^2
-2 T'h'(\del T.\del h) \nn
&&~~~~~~~~~~~~~~~~+f^{-1}u^{-3}((\del T)^2(\del h)^2-(\del T.\del h)^2)
\biggr) \biggr], \nn
\frac{1}{2} \frac{\del \Delta_T}{\del h'} &=& \frac{fu^{3/2}}{4}h'+
\frac{fh'}{4Q}(\del T)^2-\frac{fT'}{4Q} \del T.\del h, \nn
\frac{1}{2} \frac{\del \Delta_T}{\del(\del_\mu h)} &=& d_T \beta_2 \del^\mu h+
d_T \beta_3 \del^\mu T+\frac{u^{-3}}{4Q}\biggl(\del^\mu h(\del T)^2-
\del^\mu T(\partial h.\partial T) \biggr), \nn
\frac{1}{2} \frac{\del \Delta_T}{\del h} &=& T^2h \biggl[1-\frac{f^2u^{3/2}}
{4Q^2} \biggl(h^{\prime 2} (\del T)^2+T^{\prime 2} (\del h)^2
-2 T'h'(\del T.\del h) \nn
&&~~~~~~~~~~~~~~~~+f^{-1}u^{-3}((\del T)^2(\del h)^2-(\del T.\del h)^2)
\biggr) \biggr]. 
\label{delvar}
\eea
Using these one can show that
\bea
\Delta_T-T'\frac{1}{2} \frac{\del \Delta_T}{\del T'}-\del_\mu T
\frac{1}{2} \frac{\del \Delta_T}{\del(\del_\mu T)}
&=& d_T-T^{\prime 2}+\frac{u^{-3/2}}{4}(\del h)^2, \nn
T'\frac{1}{2} \frac{\del \Delta_T}{\del h'}+\del_\mu T
\frac{1}{2} \frac{\del \Delta_T}{\del(\del_\mu h)}
&=& \frac{fu^{3/2}}{4}T'h'+\frac{u^{-3/2}}{4}(\del T.\del h).
\label{delvar2}
\eea
We can now give the equations of motion obtained from the action
\eq{xuaction1}:
\bea
&& \frac{u^{13/4}}{\sqrt{\Delta_T}}\biggl[
\frac{1}{2} \frac{\del \Delta_T}{\del T}+\frac{V'}{V}\biggl(d_T-T^{\prime 2}+
\frac{u^{-3/2}}{4}(\del h)^2
\biggr)\biggr] \nn
&&~~~~~~~~~~~~~~~~~~~~~~~~~~~~= \biggl(\frac{u^{13/4}}{\sqrt{\Delta_T}}
\frac{1}{2} \frac{\del \Delta_T}{\del T'}\biggr)'+
\del_\mu \biggl(\frac{u^{13/4}}{\sqrt{\Delta_T}}
\frac{1}{2} \frac{\del \Delta_T}{\del(\del_\mu T)}\biggr), 
\label{xut} \\
&& \frac{u^{13/4}}{\sqrt{\Delta_T}}\biggl[
\frac{1}{2} \frac{\del \Delta_T}{\del h}-\frac{V'}{V}\biggl(
\frac{fu^{3/2}}{4}T'h'+\frac{u^{-3/2}}{4}(\del T.\del h)\biggr)\biggr] \nn
&&~~~~~~~~~~~~~~~~~~~~~~~~~~~~= \biggl(\frac{u^{13/4}}{\sqrt{\Delta_T}}
\frac{1}{2} \frac{\del \Delta_T}{\del h'}\biggr)'+
\del_\mu \biggl(\frac{u^{13/4}}{\sqrt{\Delta_T}}
\frac{1}{2} \frac{\del \Delta_T}{\del(\del_\mu h)}\biggr).
\label{xuh}
\eea
These can be further simplified using the expressions given in
\eq{delvar}, but we will not do so here since we will only be 
interested in a leading solution to these equations in the limit $u
\sim u_0$. As a check, we note that these equations reduce to the 
equations \eq{eq-t} and \eq{eq-l} if $T$ and $h$ are $x$-independent.

\newcommand{\sbibitem}[1]{\bibitem{#1}}

\end{document}